\documentclass[12pt]{iopart}

\usepackage{epsfig}
\usepackage{ifpdf}
\usepackage{graphicx}

\usepackage{amssymb,lineno,mathbbol,upgreek}

\usepackage[tight]{subfigure}

\usepackage{mathrsfs}

\usepackage{bbm}
\usepackage{slashed}
\usepackage{calligra}
\DeclareMathAlphabet{\mathcalligra}{T1}{calligra}{m}{n}
\DeclareFontShape{T1}{calligra}{m}{n}{<->s*[2.2]callig15}{}

\usepackage{calligra}
\DeclareMathAlphabet{\mathcalligra}{T1}{calligra}{m}{n}
\DeclareFontShape{T1}{calligra}{m}{n}{<->s*[2.2]callig15}{}
\usepackage{enumitem}

\begin{document}

\title[Relativistic soliton stability analysis]{The nonlinear Dirac equation in Bose-Einstein condensates: II. Relativistic soliton stability analysis}

\author{L H Haddad$^1$ and Lincoln D Carr$^{1,2}$}
\address{$^1$Department of Physics, Colorado School of Mines, Golden, CO 80401,USA  \\$^2$Physikalisches Institut, Universit\"at Heidelberg, D-69120 Heidelberg, Germany}

\ead{\mailto{laith.haddad@gmail.com}, \mailto{lcarr@mines.edu}}

\begin{abstract}
The nonlinear Dirac equation for Bose-Einstein condensates in honeycomb optical lattices gives rise to relativistic multi-component bright and dark soliton solutions. Using the relativistic linear stability equations, the relativistic generalization of the Boguliubov-de Gennes equations, we compute soliton lifetimes against quantum fluctuations and classify the different excitation types. For a Bose-Einstein condensate of $^{87}\mathrm{Rb}$ atoms, we find that our soliton solutions are stable on time scales relevant to experiments. Excitations in the bulk region far from the core of a soliton and bound states in the core are classified as either spin waves or as a Nambu-Goldstone mode. Thus, solitons are topologically distinct pseudospin-$1/2$ domain walls between polarized regions of $S_z = \pm 1/2$. Numerical analysis in the presence of a harmonic trap potential reveals a discrete spectrum reflecting the number of bright soliton peaks or dark soliton notches in the condensate background. For each quantized mode the chemical potential versus nonlinearity exhibits two distinct power law regimes corresponding to the free-particle (weakly nonlinear) and soliton (strongly nonlinear) limits. 
\end{abstract}

\submitto{\NJP}
\maketitle

\section{Introduction}
\label{introduction}

Vacuum states with broken symmetry play an important role in the study of quantum many-body physics, since they provide clues to the principles that govern the full symmetric theory~\cite{Goldstone1961,1nambu61,2nambu61}. Solitons are finite energy solutions of classical equations of motion and have been studied as nonuniform ground states, i.e., bound states or defects in the fundamental degrees of freedom that provide a launching point for perturbative expansions. Broken translational, rotational, or inversion symmetry, ubiquitous to discrete as well as continuous systems, can usually be cast in terms of a topological framework~\cite{Nelson2002}. When attractive interactions are present non-topological solitons model globally regular bound states of the system~\cite{Axenides2000,Diaz2007}. Such states owe their existence to an unbroken symmetry of the Lagrangian and thus have a conserved Noether charge. In contrast, topological solitons are defects typically associated with spontaneous symmetry breaking. In this case the defect breaks a discrete symmetry and appears as a boundary separating two degenerate asymptotically flat solutions while retaining a topological charge degree of freedom as a relic of the broken symmetry. Examples of solitons in extant physical systems include domain walls in BCS superconductors~\cite{Houzet2006}, superfluid vortices~\cite{London1938,Landau1941,Onsager1949}, and quantum Hall states in topological insulators~\cite{Laughlin1981,Kane2005}. In one spatial dimension dark~\cite{Lewenstein1999,Denschlag2000,Weller2008} and bright~\cite{Khaykovich2002,Carr2002,Reatto2002,Salasnich2004} solitons in repulsive or attractive Bose-Einstein condensates (BEC) with spontaneously broken U(1) symmetry are examples of broken spatial symmetry. Beyond familiar condensed matter systems solitons emerge in low-energy sectors of the standard model of particle physics as extended particles~\cite{Finkelstein1956,Enz1963,Soler1970}, and in M-theory as subcritical dimensional D-brane embeddings~\cite{Kaku1999}.

In all of these cases, one is typically interested in the properties of the low-energy spectrum since this characterizes the system near equilibrium. The presence of a defect, or soliton, partitions the domain into a core region which spans the size of the defect, and a bulk region far from the core. Excitations in the bulk describe the system's response to the presence of the soliton, whereas fluctuations in the core describe undulations and translations of the soliton itself. In superfluid systems, soliton core bound states may be metastable, possessing a finite lifetime against dissipation through lower energy scattering states, or truly stable if the soliton lies at an energy minimum of the system.

\begin{figure}[h]
\centering
 \subfigure{
\label{fig:ex3-a}
\includegraphics[width=.9\textwidth]{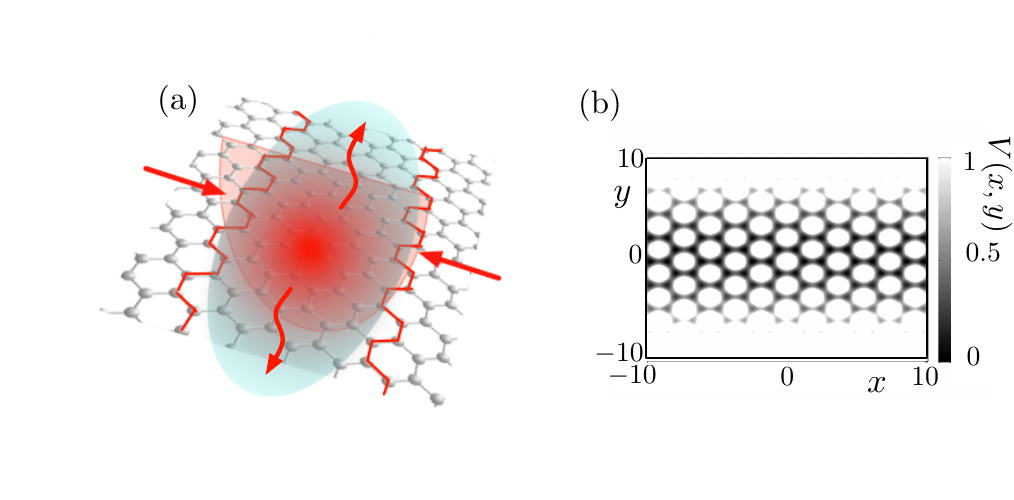} }\\
\caption[]{\emph{A soliton in the dimensionally reduced honeycomb optical lattice}. (a) Depiction of a soliton in the armchair reduction of the honeycomb optical lattice. The deep red center represents either a dark or bright soliton with fluctuations along the direction of the soliton depicted as curved arrows. The straight arrows indicate the planar direction for the quasi-1D confinement. (b) Harmonic confining potential parallel to the plane of the lattice producing the armchair pattern.} 
\label{OneD}
\end{figure}

In this article we focus on elementary excitations and stability of a topological defect near the Dirac point of a BEC. At very low temperatures interactions between condensate and non-condensate atoms is minimal, allowing for existence of long-lived metastable states. Thus, quantum fluctuations of a kink-like soliton in the nonlinear Dirac equation (NLDE) presents an analog of a domain wall in a gas of Dirac fermions interacting through a local quartic term~\cite{Watabe2012}. Solution profiles for the soliton backgrounds were explored analytically and numerically in a companion paper~\cite{haddadcarrsoliton1}. We note that similar solitons appear in nonlinear optics~\cite{ablowitz2009,Peleg2007}, in graphene~\cite{Efremidis2003,Bahat2008,Kartashov2011,Segev2010,Christodoulides1988,Segev2000}, and in various other fields of physics~\cite{Park2009,Block2010,Szameit2011,Bahat2010,Dellar2011,Ablowitz2010,CB2010,Chen2011,Kapit2011,Zhang2009,Gupta2010}. Figure~\ref{OneD} provides a schematic of our setup depicting a soliton and its fluctuations in the quasi-one-dimensional (quasi-1D) reduction of the honeycomb lattice to the armchair nanoribbon. The solution of the relativistic linear stability equations (RLSE) gives us the linear spectrum from the presence of small quantum fluctuations in the BEC~\cite{haddad2011,Haddad2012}. For both dark and bright soliton solutions, far from soliton core the BEC occupies only one of the two sublattices, switching from the A sublattice to the B sublattice when translating through the core. Thus these solution types present a 1D analog of a skyrmion localized to the soliton width. However, the skyrmion analogy does not hold near the core since in our case the total density $\rho(x)$ is a non-constant function of the longitudinal coordinate $x$. The total density here is defined as the sum of the squared spinor amplitudes, which in the case of the reduced two-spinor formulation is $\rho(x) \equiv |\psi_A(x)|^2 + |\psi_B(x)|^2$, with $\psi_A(x)$ and $\psi_B(x)$ the wavefunctions corresponding to A and B sublattices of the honeycomb lattice. We will show that quasi-particle excitations far from the soliton exist as scattering states which respect this asymmetry. Because of this feature, it is convenient to think of the switching point from the A to B sublattice as a defect analogous to a domain wall.

It is instructive to view the NLDE from a mathematically elegant perspective by recasting it in terms of the covariant pseudospin formalism. As we will see, this approach allows for a domain wall interpretation which connects to other areas of physics. For example, in magnetic systems domain walls appear as topologically stable solitons separating two distinct regions of different magnetic polarization~\cite{Spaldin2011}. Another context is the case of two interpenetrating BECs comprised of atoms in different hyperfine states, wherein one finds regions across which the relative phase of the two condensates changes by $2 \pi$~\cite{Son2002}. In spin-1 BECs, domain walls have been studied extensively as boundaries between regions of pseudospin polarization $S_z = \pm 1$~\cite{Stenger1998,Miesner1999}, in addition to investigations into the quasi-particle transmission and reflection properties of such boundaries~\cite{Watabe2012}. 

Domain walls also play an important role in high energy physics, for example as extended supersymmetric objects which isolate different vacua~\cite{Vafa1993,Abraham1992}. It is thus not surprising that solitonic objects play an important role in both condensed matter and particle physics settings. A particular example which highlights this fact is the recent simulation of tachyon condensation using two-component BECs~\cite{Takeuchi2012}. In such analogs one finds that spontaneous symmetry breaking occurs in a two-dimensional subspace of the full system, i.e., a domain wall in the larger space. In each of the examples mentioned here the domain wall is identified with a continuous deformation of the order parameter between two degenerate asymptotically flat states of the system. The key feature of the deformation is that it is localized; it occurs over a finite region in at least one of the spatial dimensions.

This article is organized as follows. Section~\ref{Symmetries} establishes the full symmetry of the quasi-1D NLDE order parameter manifold. This describes the set of possible order parameters determined by a series of symmetry breaking reductions from the full (3+1)-dimensional Poincar\'e group. In Sec.~\ref{SolitonStability}, we solve the RLSE numerically to determine soliton lifetimes. In Sec.~\ref{BoundFluctuations}, we solve the RLSE analytically through a method of decoupling and derive the phase and density fluctuations in the soliton core region, which include the Nambu-Goldstone mode responsible for $\mathrm{U}(1)$ symmetry breaking, i.e., Bose condensation. In Sec.~\ref{continuousspectrum}, we solve for the continuous spectrum far from the soliton core where we find the Nambu-Goldstone mode and a spin wave, the later corresponding to nonzero density fluctuations. The asymptotic spectrum naturally leads into Sec.~\ref{Pseudospin} where we formulate relativistic solitons in the language of spin-1/2 domain walls. In Sec.~\ref{QuadSpinWaves}, we analyze quantum fluctuations in light of the domain wall interpretation. In Sec.~\ref{SolitonSpectra}, we treat the spectral theory of a BEC in a weak harmonic trap. Finally, in Sec.~\ref{Conclusion} we conclude.

\section{Symmetries of the order parameter manifold}
\label{Symmetries}

The order parameter that we study is analogous to metastable vacua in high energy systems with quasi-particles and thermal excitations playing the role of virtual and real particles, respectively. Clarifying the underlying symmetries of the order parameter manifold is key towards identifying the various excitations associated with continuous symmetry breaking. In the quasi-1D NLDE~\cite{haddadcarrsoliton1}, the order parameter manifold comes from a series of symmetry breaking steps. To see this, we begin by noting that non-interacting bosons at the Dirac point of a quasi-2D honeycomb lattice occupy single-particle states in one-to-one correspondence with massless Dirac states. The $2   \times   2$ unit and Pauli matrices $\mathbb{1}, \, \sigma_x, \, \sigma_y$ are the group generators in 2D consistent with the spin and momentum vector coupled Dirac Hamiltonian $\mathcal{H}_p = c_l {  \bf \sigma }  \cdot \textbf{p}$. One may think of the absence of the third Pauli matrix $\sigma_z$ a consequence of projecting the full SU(2) group onto the coordinate plane thereby removing one degree of freedom through the reduction $\mathrm{SU}(2) \to \mathrm{U}(1) \otimes \mathrm{Spin}(2)$. Here the factor of U(1) accounts for an overall phase and the spin group Spin(2) is isomorphic to a double covering of U(1), i.e., expressed in terms of the fundamental group $\pi_1\!  \left(\mathrm{Spin}(2)\right)   \cong 2 \mathbb{Z} \cong 2  \pi_1\!  \left(\mathrm{U}(1)\right) $. This can be summarized in a short exact sequence by recalling the isomorphisms $\mathrm{U}(1) \cong \mathrm{SO}(2)$, $2 \mathbb{Z}  \cong  \mathbb{Z}_2$, then
\begin{eqnarray}
1 \, \to \, \mathbb{Z}_2 \, \to \, \mathrm{Spin(2)} \, \to \, \mathrm{SO}(2) \, \to \, 1
\end{eqnarray}
from which we write $\mathrm{Spin(2)}/ \mathbb{Z}_2  \cong  \mathrm{SO}(2)$, or equivalently $\mathrm{Spin(2)} \cong  \mathbb{Z}_2  \otimes  \mathrm{SO}(2)$. The quasi-1D theory then demands a second coordinate reduction which breaks the 2D rotation group into its reflection subgroup along either of two orthogonal directions $\mathrm{SO}(2) \to \mathbb{Z}_2 \oplus \mathbb{Z}_2$, where the two copies of $\mathbb{Z}_2$ are the reflection subgroups associated with the two orthogonal complex and real forms of the Dirac operator~\cite{haddadcarrsoliton1}. From this we see that the full symmetry of the quasi-1D NLDE order parameter manifold is 
\begin{eqnarray}
\mathcal{G}_\mathrm{NLDE}(1+1) = \mathrm{U}(1) \otimes  \mathbb{Z}_2 \otimes  \left( \mathbb{Z}_2 \oplus   \mathbb{Z}_2 \right)  \, , \label{fullsymmetry}
\end{eqnarray}
 To make this discussion more concrete, we can write the representation of this symmetry reduction from 2D to 1D in terms of the order parameter manifold as  
\begin{eqnarray}
\hspace{-4pc}  e^{i \phi }    \left( \! \begin{array}{c} 
                          e^{ - i  \theta({\bf p}) /2}  \\
                          \pm \,  e^{i  \theta({\bf p})  /2} \end{array} \! \right)   \; \; \to \;  \;    e^{i \phi }       \left( \! \begin{array}{c} 
                           i^{ ( 1 - \mathrm{ {\bf p} }/|\mathrm{ {\bf p} }|) /2}   \\
                          \pm \,  (-i)^{ ( 1 - \mathrm{ {\bf p} }/|\mathrm{ {\bf p} }|) /2}  \end{array} \! \right) \;  \;  \oplus \;   \; e^{i \phi }       \left( \! \begin{array}{c} 
                           i^{ ( 1 - \mathrm{ {\bf p}}/|\mathrm{ {\bf p}}|) /2}   \\
                          \pm \,  i \, (-i)^{ ( 1 - \mathrm{ {\bf p}}/|\mathrm{ {\bf p}}|) /2}  \end{array} \! \right) \,    \label{reduction}
\end{eqnarray}
where $\theta({\bf p}) \equiv \mathrm{tan}^{-1}(p_y /p_x)$, and in terms of the Hilbert space the reduction in Eq.~(\ref{reduction}) acts according to $H_\mathrm{2D} \, \to \,  H_x \, {\bigoplus} \,  H_y$. Here the subscripts refer to the Hilbert spaces associated with the independent 1D Dirac operators obtained by decomposing the 2D operator along two orthogonal directions in the plane: $D  = - i \hbar c_l \left(   \sigma_x \partial_x  +   \sigma_y \partial_y \right) \equiv D_x + D_y$. Note that on the left side of Eq.~(\ref{reduction}) the vector $\mathrm{ {\bf p} }$ is two-dimensional, whereas the right hand side applies to one spatial dimension. In the reduced space, the direction of $\mathrm{ {\bf p} }$ is completely determined by a sign, i.e., $\mathrm{ {\bf p} } = \pm |\mathrm{ {\bf p} }| \equiv \pm \, p$. We adhere to this convention throughout our work.

The first order parameter manifold in Eq.~(\ref{reduction}) has the full $\mathrm{U}(1) \otimes \mathrm{Spin}(2)$, where $\phi$ and $\theta$ are the U(1) and Spin(2) parameters. To the right of the arrow in Eq.~(\ref{reduction}) the order parameter takes on the reduced symmetry where $\phi$ is the U(1) parameter with the positive/negative eigenvalues and parity reversing factors associated with the $\mathbb{Z}_2 \otimes  \mathbb{Z}_2$ products in Eq.~(\ref{fullsymmetry}). 

The presence of a soliton in the reduced quasi-1D problem breaks translational symmetry, in which case one would expect to find one zero-energy mode in addition to one massless excitation for each broken continuous symmetry. These include two Goldstone modes, one from condensation in the overall phase and one from the internal phase; and two zero modes, one from breaking rotational symmetry when going from 2D to 1D, and one from the broken translational symmetry due to the soliton. Only two out of the four are in fact present. The Goldstone and zero modes from breaking Spin(2) symmetry are suppressed, since they fluctuate along the direction of the quasi-1D confining potential. We expect therefore to find one Goldstone mode as an overall phase fluctuation and a zero mode from the soliton. It must be noted that in the literature the Goldstone mode is sometimes identified as a zero mode. Technically, the Goldstone mode corresponds to the gapless energetic branch associated with local twists in the phase. When the condensate background is spatially uniform, the Goldstone branch is continuous and connects to a spatially uniform zero mode. In the presence of a defect, however, translational symmetry is broken and the Goldstone branch is discrete with a nonzero momentum lower bound, $p  \ge p_\mathrm{min}$. In this case the Goldstone branch connects to a spatially nontrivial zero mode in the limit $p  \to p_\mathrm{min}$.

The nonlinearity in the NLDE allows for asymptotically flat solutions $|\psi_A|, \, |\psi_B|  \to 0 ,  \, \sqrt{\mu/U}$, for $|x|$ much larger than the soliton core size. The $\mathbb{Z}_2 \otimes  \mathbb{Z}_2$ symmetry in Eq.~(\ref{reduction}) leads to four distinct asymptotic states but only two are independent because of an overall phase redundancy. These are
\begin{eqnarray}
  \left( \! \begin{array}{c} 
                           1   \\
                          +  1   \end{array} \! \right) \; , \;\;\;  \left( \! \begin{array}{c} 
                           1   \\
                            - 1   \end{array} \! \right) \; ,   \label{zigzagvacuua}
\end{eqnarray}
for the Dirac operator $D_y$, and 
\begin{eqnarray}
  \left( \! \begin{array}{c} 
                           1   \\
                          +  i    \end{array} \! \right) \; , \;\;\;  \left( \! \begin{array}{c} 
                           1   \\
                            - i    \end{array} \! \right) \; ,  \label{armchairvacuua}
\end{eqnarray}
for $D_x$, with an overall complex constant prefactor omitted for clarity. As we showed in~\cite{haddadcarrsoliton1}, NLDE solitons interpolate between two asymptotic states that are linear combinations of
\begin{eqnarray}
  \left( \! \begin{array}{c} 
                           1   \\
                           0    \end{array} \! \right) \; , \;\;\;  \left( \! \begin{array}{c} 
                            0   \\
                             1  \end{array} \! \right) \; ,   \label{zigzagvacuua}
\end{eqnarray}
associated with $D_y$, and 
\begin{eqnarray}
  \left( \! \begin{array}{c} 
                           1   \\
                           0    \end{array} \! \right) \; , \;\;\;  \left( \! \begin{array}{c} 
                           0    \\
                             i    \end{array} \! \right) \; ,  \label{armchairvacuua}
\end{eqnarray}
associated with $D_x$. We will see in this article that the presence of a soliton partially breaks the inversion symmetry implicit in Eqs.~(\ref{zigzagvacuua})-(\ref{armchairvacuua}), splitting the spectrum into massless modes with linear dispersion, which retain the full symmetry, and massive modes with quartic dispersion, which break parity inversion symmetry. The central focus of this article is to understand the nature of these quantum fluctuations, both asymptotically and in the transition region inside the soliton core.

\section{Stability of soliton solutions}
\label{SolitonStability}
 
% \begin{eqnarray}
% U_\mathrm{1D} &\equiv  U_\mathrm{2D}  \left( \frac{M \omega_y}{ \pi \hbar}\right)^{1/4} \hspace{-.5pc}   \int_{-\infty}^{+ \infty} \hspace{-.5pc} \! dy \,  \mathrm{exp}\left( - 2 M \omega_y y^2/ \hbar \right)   =   \frac{U_\mathrm{2D}}{ \pi^{1/2} L_y}      \, , 
%\end{eqnarray}

The combination of the honeycomb lattice geometry and the atom-atom interactions results in a characteristic signature effect on soliton stabilities. In particular, the presence of negative energy states below the Dirac point means that a BEC will eventually decay by radiating into the continuum of negative energy scattering states. However, this requires a mechanism for energy dissipation into non-condensate modes which must come about from secondary interactions with thermal atoms. Thus, as long as the system is at very low temperatures our main concern for depletion of the BEC comes from potential imaginary eigenvalues in the linear spectrum. The situation is analogous to dark solitons in quasi-1D BECs described by the nonlinear Schr\"odinger equation: in practice such excited states can easily have a lifetime longer than that of the BEC~\cite{stellmerS2008}. In this section, we compute the linear spectrum for soliton solutions of the quasi-1D NLDE.

 Before proceeding it is useful to elaborate on units and dimensions of some of the physical quantitates key to our discussion. The main composite parameters relevant to the NLDE, and hence the RLSE, are the effective speed of light $c_l = t_h a \sqrt{3}/2 \hbar$ and the quasi-1D renormalized atom-atom binary interaction strength $U_\mathrm{1D} =   U_\mathrm{2D}/(\pi^{1/2} L_y)$, expressed in terms of its quasi-2D counterpart $U_\mathrm{2D} =  L_z g \bar{n}^2 3 \sqrt{3} a^2/8$. The presence here of the trap oscillator lengths, $L_y$ and $L_z$, reflect the fact that $U_\mathrm{1D}$ and $U_\mathrm{2D}$ come from integrating over the degrees of freedom transverse to the single large dimension in our problem. For instance, $U_\mathrm{1D}$ is obtained by integrating over the ground state in the $y$-direction in the quasi-2D NLDE~\cite{Haddad2012} 
\begin{eqnarray}
 U_\mathrm{1D} \equiv  U_\mathrm{2D}   \left( \frac{3  }{2 L_y }\right)^{2} \hspace{-.5pc}   \int_{- L_y/2}^{+  L_y/2 } \hspace{-.5pc} \! dy \, \left(   1  - 4  \frac{y^2}{L_y^2 } \right)   =   \left( \frac{6 }{ 5 L_y }\right)  U_\mathrm{2D}   \, , 
\end{eqnarray}
where the oscillator length is related to the frequency $\omega_y$ and atomic mass $M$ by $L_y = \sqrt{\hbar /M \omega_y}$. The parameters that comprise $U_\mathrm{2D}$ and $c_l$ are the vertical oscillator length $L_z$ (in the quasi-2D problem), the average particle density $\bar{n} = N/V$, the interaction $g= 4 \pi \hbar^2 a_s/M$, the lattice constant $a$, and the hopping energy $t_h$. Throughout our work we take the atomic mass $M$ and scattering length $a_s = 5.77 \, \mathrm{nm}$ to be those of $^{87}\mathrm{Rb}$. A complete discussion of NLDE parameters and constraints can be found in~\cite{Haddad2012}. With these parameter definitions one finds that the spinor order parameter $\Psi = \left(\psi_A, \, \psi_B\right)$ is dimensionless and the quasi-1D interaction strength $U_\mathrm{1D}$ has dimensions of energy. To simplify the notation, from here on we will omit the subscript on $U_\mathrm{1D}$ and write $U$ for the quasi-1D interaction strength.

To compute soliton lifetimes we must solve the relativistic linear stability equations (RLSE) modified for our quasi-one-dimensional problem~\cite{haddad2009}. This allows us to account for quantum mechanical perturbations to the mean-field result by using the corrected order parameter
\begin{eqnarray}
\hat{\psi}(x, t) = e^{-i \mu  t /\hbar } \left[  \,  \Psi(x) + \hat{\phi}(x, t)\,  \right]\, ,  \label{psisplit1}
\end{eqnarray}
with the condensate spinor wavefunction and quantum correction given by 
\begin{eqnarray}
\Psi(x)  =   \left[ \, \psi_A(x) , \, \psi_B(x) \, \right]^T , \label{psisplit2} \\
 \hat{\phi}(x, t ) =   e^{ - i E t/\hbar }  \left[ \,   \hat{\alpha} \, u_A(x) , \, \hat{\beta} \, u_B(x) \, \right]^T  -  e^{  i E t/\hbar }   \left[ \, \hat{\alpha}^\dagger v_A^*(x) , \, \hat{\beta}^\dagger v_B^*(x)\,  \right]^T   ,  \label{psisplit3}
\end{eqnarray}
where $\hat{\alpha}^\dagger$ and $\hat{\beta}^\dagger$ ($\hat{\alpha}$ and $\hat{\beta}$) are the creation (destruction) quasi-particle operators and $u_{A(B)}$ and $v_{A(B)}$ are the associated spatial functions, respectively. Linear stability of a particular soliton solution is determined by substituting the spatial function for that solution (i.e., the dark or bright soliton) into the RLSE as a background for the quasi-particle functions. This substitution gives a set of first-order coupled ODEs in one independent variable to be solved consistently for the quasi-particle energies $E_k$ and amplitudes $\mathrm{ \bf u}_\mathrm{ \bf k}$ and $\mathrm{ \bf v}_\mathrm{ \bf k}$, where the subscript denotes the mode with momentum $p = \hbar |\mathrm{ \bf k}|$, defined in terms of the magnitude of the wavevector $\textbf{\textrm{k}}$. We remind the reader that we are working in one spatial dimension, thus there is at most a sign difference between the bold vector notation and the corresponding norm: $\textbf{\textrm{k}} = \pm | \textbf{\textrm{k}}|$. Since we are perturbing from a spin-$1/2$ BEC background, $\mathrm{ \bf u}_\mathrm{ \bf k}(x) = \left[ u_{k A }(x), \, u_{k B }(x) \right]^T$ and $\mathrm{ \bf v}_\mathrm{ \bf k}(x) = \left[ v_{ k A}(x), \, v_{ k B}(x) \right]^T$ have vector form describing quasi-particle and quasi-hole excitations of the A and B sublattice, as indicated by the A(B) sublattice subscripts. We discretize the derivatives and spatial functions in the RLSE using a forward-backward average finite-difference scheme, then solve the resulting discrete matrix eigenvalue problem using the Matlab function {\it eig}. 

Solutions of the RLSE are perturbations of the NLDE four-spinor components and respect the same decoupling to two-spinor form. Thus, focusing on equations for the upper two-spinor, the 1D RLSE is 
\begin{eqnarray}
\fl  - \hbar  c_l \frac{\partial u_{k B}}{\partial x}   - \mu  u_{k A} +2  U \left| \psi_A \right|^2 u_{k A} - U \left| \psi_A \right|^2 v_{k A}  &= E_k  u_{k A} \, ,  \label{eqn:RLSE1}\\
 \fl    \hbar  c_l     \frac{\partial u_{k A}}{\partial x}- \mu  u_{k B} + 2 U \left| \psi_B \right|^2 u_{k B} - U\, \left| \psi_B \right|^2  v_{ k B}  &= E_k   u_{k B}  \, ,  \label{eqn:RLSE2}    \\
  \fl \hbar c_l  \frac{\partial v_{k B}}{\partial x}-\mu v_{ k A} + 2U \left| \psi_A \right|^2  v_{ k A} -  U \left| \psi_A \right|^2 \, u_{k A}  &= - E_k   v_{ k  A}\, ,     \label{eqn:RLSE3} \\
\fl -  \hbar  c_l  \frac{\partial v_{k A}}{\partial x}- \mu v_{k B} + 2U \left| \psi_B \right|^2 v_{k B} -  U\, \left| \psi_B \right|^2 u_{k B}  &=  - E_k   v_{k B} \, .  \label{eqn:RLSE4} 
\end{eqnarray}
Equations~(\ref{eqn:RLSE1})-(\ref{eqn:RLSE4}) inherit the linear derivative structure on the sublattice particle and hole functions $u_{A(B)}$ and $v_{A(B)}$. The constant chemical potential $\mu$ and particle interaction $U$ appear as coefficients in addition to the spatially dependent condensate profiles $\psi_{A(B)}(x)$ and eigenvalues $E_k$. The parameters in Eqs.~(\ref{eqn:RLSE1})-(\ref{eqn:RLSE4}) are already renormalized due to dimensional reduction from 2D to quasi-1D as described in Sec.~3 of Ref.~\cite{haddadcarrsoliton1}. We point out that Eqs.~(\ref{eqn:RLSE1})-(\ref{eqn:RLSE4}) pertain to the NLDE associated with the real Dirac operator. In the complex version the momentum terms have identical complex coefficients, $-i \hbar c_l$, which comes from rotating the Dirac operator by $90$ degrees. This transformation between real and complex forms is equivalent to the two-spinor Pauli transformation discussed in Sec.~2 in Ref.~\cite{haddadcarrsoliton1}, and the four equations of the RLSE inherit this feature: choosing to work in one form leads to no loss of generality. Alternatively, one may argue that since the RLSE are linear in the amplitudes $u_{A(B)}$ and $v_{A(B)}$, absorbing a factor of $i$ into either pair of the sublattice amplitudes, i.e., $u_{A}$ and $v_{A}$ or $u_{B}$ and $v_{B}$, simply converts between the real and complex forms. Thus, for a given condensate spatial profile the RLSE for the real and complex Dirac operator have the same linear eigenvalues, and the stability properties of solitons in both cases are the same.

\begin{figure}[h]
\centering
 \subfigure{
\label{fig:ex3-a}
\includegraphics[width=6in,height=2.2in]{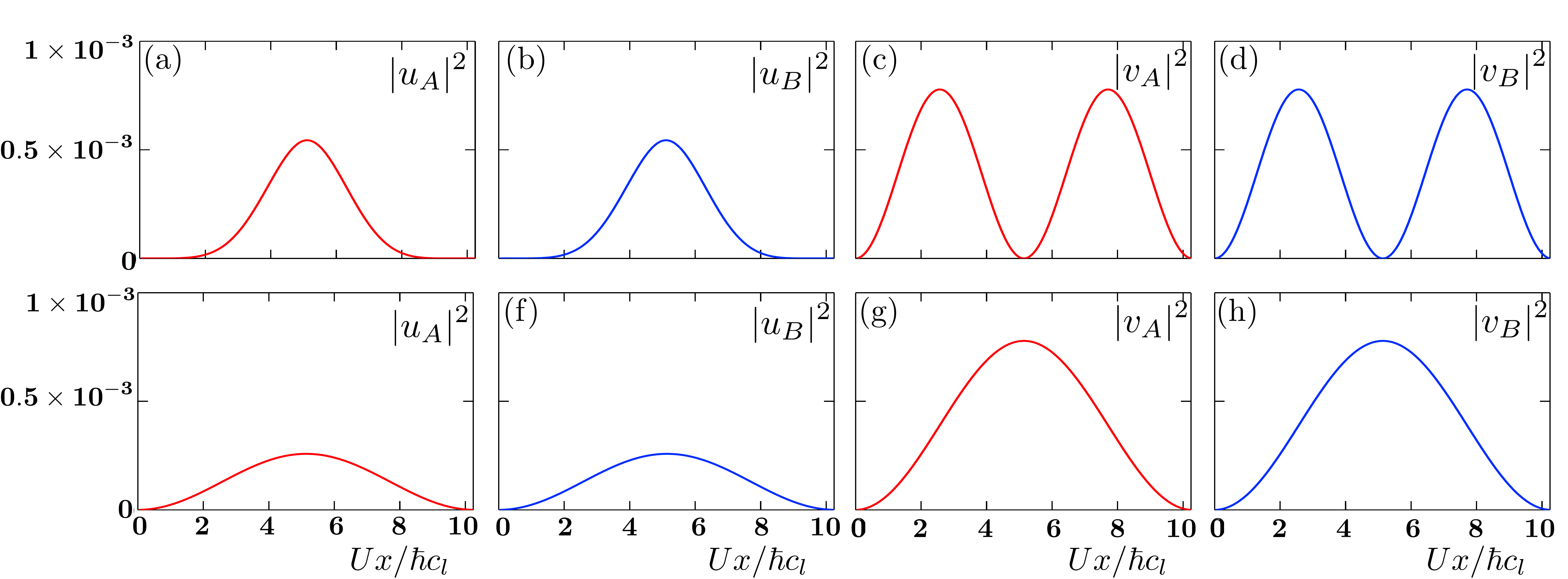}}\\ 
\caption[]{\emph{Soliton quasi-particle excitations in the quasi-1D reduction of the NLDE}. (a)-(d) Excitations of the dark soliton near the defect (core) of the soliton. (e)-(f) Excitations of the bright soliton. These excitations are real, up to a constant phase factor, in contrast to scattering states far from the center of the soliton. } \label{Quasiparticles}
\end{figure}

We find the lowest excitation energies for the two types of solitons $E_1^\textrm{DS} = \pm 0.1862 \,U$ and $E_1^\textrm{BS} =  \pm  0.1902 \,U$, in units of the interaction $U$, where the superscripts DS and BS refer to the dark and bright solitons, respectively, for the quasi-1D NLDE. Figure~\ref{Quasiparticles} shows the associated quasi-particle functions which are bound states at the defect point of the soliton, i.e., near the region where the density transitions from the A to the B sublattice. The bound states shown in Fig.~\ref{Quasiparticles} decay far from the soliton core where the continuum of scattering states is dominant. The negative eigenvalues correspond to modes which decrease the energy of the solitons into states below the Dirac point. What is significant is the absence of imaginary modes; thus our solitons are dynamically stable. This means that at very low temperatures we expect solitons to remain viable over the lifetime of the BEC. To obtain the next order correction due to finite temperature effects would require a modified version for the RLSE analogous to the Hartree-Fock-Bogoliubov treatment which takes into account interactions between condensate and non-condensate atoms~\cite{Snyder2012}.

\section{Bound state fluctuations of the soliton core}
\label{BoundFluctuations}

We would like to solve for the quasi-particle structure of the NLDE using analytical methods. Towards this end, in this section we reduce the RLSE down to four decoupled second order equations. We begin by changing variables using symmetric and antisymmetric functions defined as 
\begin{eqnarray}
 \psi_+  \equiv \frac{1}{2} \left( | \psi_A |^2 + |\psi_B|^2 \right) , \;\; \psi_-  \equiv  \frac{1}{2} \left( | \psi_A|^2 - |\psi_B|^2 \right) ,\label{transf1} \\
 u_+ \equiv \frac{1}{2} \left( u_A  + u_B  \right) , \;\; u_-  \equiv  \frac{1}{2} \left(  u_A  - u_B  \right) ,\label{transf2} \\
 v_+ \equiv \frac{1}{2} \left( v_A  + v_B \right) , \;\; v_-  \equiv  \frac{1}{2} \left(  v_A -  v_B  \right) ,  \label{transf3}
\end{eqnarray}
where we have suppressed the mode index $k$ in order to simplify the notation. Using the transformation defined by Eqs.~(\ref{transf1})-(\ref{transf3}), Eqs.~(\ref{eqn:RLSE1})-(\ref{eqn:RLSE4}) become 
\begin{eqnarray}
\fl  \hbar c_l u_-'  +  \mu \, u_+  -  2 U \left( \psi_+ u_+ + \psi_- u_- \right) + U \left( \psi_+ v_+ + \psi_- v_- \right) &=  - E \,  u_+  \, , \\
\fl  \hbar c_l u_+'  +  \mu \, u_-   -   2 U \left( \psi_- u_+ + \psi_+ u_- \right) + U \left( \psi_- v_+ + \psi_+ v_- \right) &=  - E \, u_-  \, , \\
\fl  \hbar c_l v_-'  +  \mu \, v_+  -  2 U \left( \psi_+ v_+ + \psi_- v_- \right) +  U \left( \psi_+ u_+ + \psi_- u_- \right) &= + E \, v_+ \, , \\
\fl  \hbar c_l v_+'  +  \mu \, v_-  - 2 U \left( \psi_- v_+ + \psi_+ v_- \right) + U \left( \psi_- u_+ + \psi_+ u_- \right) &=  + E \, v_-  \, ,
\end{eqnarray}
which we obtain by adding and subtracting the transformed versions of Eqs.~(\ref{eqn:RLSE1})-(\ref{eqn:RLSE2}) and Eqs.~(\ref{eqn:RLSE3})-(\ref{eqn:RLSE4}). Note that we use the prime notation to denote derivatives with respect to $x$. These equations lead to a further decoupling into two pairs of equations. Transforming to the dimensionless form using the new variable and constants $\tilde{\mu} = \mu/U$, $\tilde{E}= E/U$, $\tilde{x} = Ux/\hbar c_l$, we obtain \begin{eqnarray}
 u_1' + [ \tilde{\mu} - 2 (\psi_+ + \psi_- ) + \tilde{E} ] u_1 + (\psi_+ + \psi_-) v_1 =  0 \,  , \label{dec1}\\
v_1' + [ \tilde{\mu} - 2 (\psi_+ + \psi_- ) - \tilde{E} ] v_1 + (\psi_+ + \psi_-) u_1   = 0  \, , \label{dec2}\\
u_2' + [ \tilde{\mu} - 2 (\psi_+ - \psi_- ) + \tilde{E} ] u_2 + (\psi_+ - \psi_-) v_2 =  0 \, ,\label{dec3} \\
v_2' + [ \tilde{\mu} - 2 (\psi_+ - \psi_- ) - \tilde{E} ] v_2 + (\psi_+ - \psi_-) u_2    = 0 \, , \label{dec4}
\end{eqnarray}
where we have defined new variables $u_1 = u_+ + u_-$, $v_1 = v_+ + v_-$, $u_2 = u_+ -  u_-$, $v_2  = v_+ - v_-$. Note that now Eqs.~(\ref{dec1})-(\ref{dec2}) are not coupled to Eqs.~(\ref{dec3})-(\ref{dec4}). Next, we make the substitution
\begin{eqnarray}
u_1  =  e^{- \int \! d\tilde{x} \, [ \tilde{\mu} -2(\psi_+ + \psi_-) + \tilde{E} ] }  \, \tilde{u}_1 \, , \label{decans1}  \\
v_1  =  e^{- \int \! d\tilde{x} \, [ \tilde{\mu} -2(\psi_+ + \psi_-) - \tilde{E} ] }  \, \tilde{v}_1 \, . \label{decans2}  
\end{eqnarray}
Substituting Eqs.~(\ref{decans1})-(\ref{decans2}) into Eqs.~(\ref{dec1})-(\ref{dec2}), we obtain
\begin{eqnarray}
\tilde{u}_1' + (\psi_+ + \psi_-) \tilde{v}_1 e^{+ 2 \tilde{E} x}    =   0 \label{utilde} \, ,\\
\tilde{v}_1'  + (\psi_+ + \psi_-) \tilde{u}_1 e^{-2 \tilde{E} x}    =  0 \, . \label{vtilde}
\end{eqnarray}
Back substitution yields the decoupled second order equations 
\begin{eqnarray}
 \tilde{u}_1'' - \left[ 2 \tilde{E}  + \mathrm{ln}( \psi_+ + \psi_- ) ' \right]  \tilde{u}_1' - (\psi_+ + \psi_- )^2\tilde{u}_1  =   0 \,   , \label{secondorder1} \\
   \tilde{v}_1'' +  \left[  2 \tilde{E}  - \mathrm{ln}( \psi_+ + \psi_- ) ' \right]  \tilde{v}_1'  - (\psi_+ + \psi_- )^2 \tilde{v}_1  =  0 \, .  \label{secondorder2}
\end{eqnarray}
These equations may be further simplified by transforming to standard form using the substitution
\begin{eqnarray}
 \tilde{u}_1 = e^{\frac{1}{2} \!  \int \! d\tilde{x} \, [2 \tilde{E} + \mathrm{ln}(\psi_+ + \psi_-)' ] } \,  w \,  , 
 \end{eqnarray}
and 
\begin{eqnarray}
\tilde{v}_1 = e^{ - \frac{1}{2} \!  \int \! d\tilde{x} \, [2 \tilde{E}  - \mathrm{ln}(\psi_+ + \psi_-)' ] }  \,  z \,.  
\end{eqnarray}
Eq.~(\ref{dec1})-(\ref{dec2}) are reduced to 
\begin{eqnarray}
w'' + Q(\tilde{x}) \,  w =  0  \, ,\label{finalform1}\\
z'' + R(\tilde{x}) \, z =  0  \, , \label{finalform2}
\end{eqnarray}
where 
\begin{eqnarray}
 Q(\tilde{x}) =&  -(\psi_+ + \psi_-)^2 +  \frac{1}{2}  \left[ \mathrm{ln}(\psi_+ + \psi_-) \right]''   \label{Q}\\
 &- \tilde{E}^2  - \,  \tilde{E}\,   \mathrm{ln}( \psi_+ + \psi_- ) '  - \frac{1}{4} \, \mathrm{ln}^2( \psi_+ + \psi_- ) ' \, , \nonumber  \\
R(\tilde{x})  =&  -(\psi_+ + \psi_-)^2 +  \frac{1}{2}  \left[ \mathrm{ln}(\psi_+ + \psi_-) \right]''   \label{R} \\
&- \tilde{E}^2 +\,   \tilde{E}\,   \mathrm{ln}( \psi_+ + \psi_- ) '  - \frac{1}{4} \, \mathrm{ln}^2( \psi_+ + \psi_- ) ' \, .  \nonumber 
\end{eqnarray}
Following a similar path yields the reduced forms for Eqs.~(\ref{dec3})-(\ref{dec4})
\begin{eqnarray}
q'' +   S(\tilde{x}) \, q =  0  \, \label{finalform3},\\
s'' +  T(\tilde{x}) \, s =  0  \, , \label{finalform4} 
\end{eqnarray}
where 
\begin{eqnarray}
\hspace{0pc} S(\tilde{x}) =& -(\psi_+-  \psi_-)^2 +  \frac{1}{2}  \left[ \mathrm{ln}(\psi_+ - \psi_-) \right]''   \label{S}  \\
& - \tilde{E}^2 -  \,  \tilde{E}\,   \mathrm{ln}( \psi_+ - \psi_- ) ' - \frac{1}{4} \, \mathrm{ln}^2( \psi_+ - \psi_- ) ' \, , \nonumber  \\
T(\tilde{x}) &= -(\psi_+ - \psi_-)^2 +  \frac{1}{2}  \left[ \mathrm{ln}(\psi_+ -  \psi_-) \right]''  \label{T}  \\
& - \tilde{E}^2  + \,  \tilde{E}\,   \mathrm{ln}( \psi_+ - \psi_- ) '  - \frac{1}{4} \, \mathrm{ln}^2( \psi_+ - \psi_- ) '   \, . \nonumber 
\end{eqnarray}

Equations~(\ref{finalform1})-(\ref{finalform2}) and~(\ref{finalform3})-(\ref{finalform4}) comprise the final reduced form of the RLSE in terms of four decoupled second order equations. The quasi-particle amplitudes can be obtained from the functions $w$, $z$, $q$ and $s$ by working backwards through each transformation step by which we obtain 
\begin{eqnarray}
u_A  =  | \psi_A| \, e^{- \int \! d\tilde{x} \,    ( \tilde{\mu} -2 |\psi_A|^2  )    }  \, w \,  \label{uaw} , \\
v_A   =   | \psi_A| \, e^{- \int \! d\tilde{x} \, (  \tilde{\mu} -2 |\psi_A|^2 )   }  \, z \, ,\label{vaw}   \\
 u_B  =  | \psi_B| \, e^{ - \int \! d\tilde{x} \, ( \tilde{\mu} -2 |\psi_B|^2  )  }  \, q \,\label{ubq} , \\
v_B  =  | \psi_B| \, e^{ -  \int \! d\tilde{x} \, ( \tilde{\mu} -2 |\psi_B|^2  )   }  \, s \,\label{vbq}  . 
\end{eqnarray}
It is also interesting to note that Eqs.~(\ref{finalform1})-(\ref{finalform2}) and~(\ref{finalform3})-(\ref{finalform4}) have a Schr\"odinger-like form for a particle with zero total energy subject to the potential functions $Q$, $R$, $S$ and $T$. For soliton solutions the potential functions develop minima which support stable localized bound states. It is also important to note that Eqs.~(\ref{finalform1})-(\ref{finalform2}) and~(\ref{finalform3})-(\ref{finalform4}) must be solved self consistently since the potentials in Eqs.~(\ref{Q})-(\ref{R}) and Eqs.~(\ref{S})-(\ref{T}) depend on the eigenvalues $\tilde{E}$. In general, the RLSE allow for scattering states and bound states in the $u$'s and $v$'s.

For a given soliton background, bound state solutions of the RLSE describe fluctuations in the density and phase of the soliton core. Here we examine the bound state regimes by computing the density and phase fluctuations near the core using the exact reductions of the RLSE in Eqs.~(\ref{uaw})-(\ref{vbq}). From Eqs.~(\ref{psisplit1})-(\ref{psisplit3}), the fluctuations in the density and phase are given by 
\begin{eqnarray}
\fl  \delta \rho(x, t) = | \langle \hat{\psi}(x, t) \rangle |^2  -   | \Psi(x) |^2  \nonumber \\
          =      \left[  u_A(x) \psi_A^*(x)  - v_A(x) \psi_A(x)  \right] 	e^{-i E t/\hbar} \nonumber \\
       +   \left[  u_A^*(x) \psi_A(x)  - v_A^*(x) \psi_A^*(x)  \right] 	e^{+ i E t/\hbar} \nonumber  \\
        +  \left[  u_B(x) \psi_B^*(x)  - v_B(x) \psi_B(x)  \right] 	e^{-i E t/\hbar} \nonumber \\
         +   \left[  u_B^*(x) \psi_B(x)  - v_B^*(x) \psi_B^*(x)  \right] 	e^{+ i E t/\hbar} + \dots  \nonumber \\
         \hspace{6pc} \equiv \delta \rho_A(x, t) + \delta \rho_B(x, t)  + \mathcal{O}[u^2_{A(B)}, v^2_{A(B)}] \,  , \label{densityfluct}
\end{eqnarray}
and
\begin{eqnarray}
\fl  \delta \varphi(x, t) =  \mathrm{tan}^{-1} \frac{ \mathrm{Im} \left[ u_A(x) e^{-i Et/\hbar} - v_A^*(x)  e^{+ i E t/\hbar}  \right]  }{   \mathrm{Re}\left[  \psi_A(x) \right] } \nonumber \\
            + \,   \mathrm{tan}^{-1} \frac{ \mathrm{Im} \left[ u_B(x) e^{-i Et/\hbar} - v_B^*(x)  e^{+ i E t/\hbar}  \right]  }{   \mathrm{Re}\left[  \psi_B(x) \right] } + \dots  \nonumber  \\
               \hspace{6pc}  \equiv \delta \varphi_A(x, t)  + \delta \varphi_B(x, t) + \mathcal{O}[ (u/\psi)^2_{A(B)},    (v/\psi)^2_{A(B)} ] \,    , \label{phasefluct} 
\end{eqnarray}
where we have used the fact that the square modulus of a quasi-particle amplitude is much smaller than that of the condensate wavefunction.

Next, we solve the decoupled equations~(\ref{finalform1})-(\ref{finalform2}), (\ref{finalform3})-(\ref{finalform4}) using approximate forms for the dark soliton spinor components
\begin{eqnarray}
f_A(x)  \approx  (\tilde{\mu} /2) \left[ 1 - \mathrm{tanh}(\tilde{x}) \right] \, , \label{darkapprox1} \\
f_B(x)  \approx  (\tilde{\mu} /2) \left[ 1 + \mathrm{tanh}(\tilde{x}) \right]\, . \label{darkapprox2}
\end{eqnarray}
The approximate forms in Eqs.~(\ref{darkapprox1})-(\ref{darkapprox2}) can be improved upon by introducing variational parameters for the position and width of each component. To solve for the equilibrium positions and widths one then extremizes the NLDE energy functional
\begin{eqnarray}
E\left[ \Psi^\dagger , \Psi \right] = \int \! d{\bf r}  \left\{ i \hbar c_l  \Psi^\dagger {\bf \sigma } \cdot  \nabla \Psi  +  \frac{U }{2} \sum_{i=1}^2 \left[ \Psi^\dagger \frac{1}{2} \left(\mathbb{1} + \epsilon_i \sigma_z \right)   \Psi \right]^2   \right\}  \, ,    \nonumber 
\end{eqnarray}
where $\epsilon_1 = +1$ and $\epsilon_2 = - 1$, and $\Psi = (\Psi_A , \, \Psi_B )^T$. In what follows though we will use the forms Eqs.~(\ref{darkapprox1})-(\ref{darkapprox2}). An analogous calculation for the bright soliton proceeds along the same lines but using the forms $f_A(x) \approx (\tilde{\mu} /2) \left[ 1 - \mathrm{tanh}(\tilde{x}) +  \mathrm{sech}(\tilde{x})  \right]$ and $f_B(x) \approx (\tilde{\mu} /2) \left[ 1 + \mathrm{tanh}(\tilde{x}) +  \mathrm{sech}(\tilde{x})  \right]$. Substituting Eqs.~(\ref{darkapprox1})-(\ref{darkapprox2}) into Eqs.~(\ref{uaw})-(\ref{vbq}) gives the quasi-particle amplitudes 
\begin{eqnarray}
 u_A  = \frac{\sqrt{\tilde{\mu}}}{2} \left[ 1 - \mathrm{tanh}(\tilde{x}) \right] e^{ -\tilde{\mu} \{ |\tilde{x}| + \mathrm{sech}^2(\tilde{x}) [ \mathrm{tanh}(\tilde{x}) -1 ] \} } w\,  , \nonumber \\ 
v_A  = \frac{\sqrt{\tilde{\mu}}}{2}  \left[ 1 - \mathrm{tanh}(\tilde{x}) \right]  e^{- \tilde{\mu} \{ |\tilde{x} |+  \mathrm{sech}^2(\tilde{x}) [ \mathrm{tanh}(\tilde{x}) -1 ] \} } z \, , \nonumber \\
 u_B  =  \frac{\sqrt{ \tilde{\mu}}}{2}  \left[ 1 +  \mathrm{tanh}(\tilde{x}) \right]  e^{ -  \tilde{\mu} \{  |\tilde{x}| +  \mathrm{sech}^2(\tilde{x}) [ \mathrm{tanh}(\tilde{x}) + 1 ]  \} }  q \, ,  \nonumber \\ 
 v_B =  \frac{ \sqrt{\tilde{\mu}}}{2}  \left[ 1 +  \mathrm{tanh}(\tilde{x}) \right]  e^{ - \tilde{\mu} \{ |\tilde{x}| +  \mathrm{sech}^2(\tilde{x}) [ \mathrm{tanh}(\tilde{x}) + 1 ] \} } \, s \, . \nonumber 
\end{eqnarray}
The envelope functions in these expressions decay exponentially for $x \to \pm \infty$, with the specific details of the fluctuations determined by solving for the functions $w$, $z$, $q$, and $s$ subject to the particular forms of the potentials $Q$, $R$, $S$, $T$. Computing these potentials analytically, we find three regimes: (i) all potentials are negative on the interval $- \infty < x  < + \infty$ when $|\tilde{E}| < 2 - \sqrt{3}$, which lead to exponentially growing and decaying solutions; (ii) two of the potentials, $Q$ and $R$, are identically zero and two are negative on the interval $- \infty < x  < + \infty$ when $|\tilde{E}| = 2 - \sqrt{3}$; and (iii) potentials $Q$ and $R$ become positive when $2 - \sqrt{3} < |\tilde{E}| < \tilde{\mu}$, for which complex oscillating solutions exist. The upper bound $\tilde{\mu}$ is the energy of the gapped branch of the continuous spectrum at zero quasi-particle momentum ($p \to 0$) which we will discuss in Sec.~\ref{continuousspectrum}.

We can interpret the regimes (i) and (ii) in light of the density and phase fluctuations, Eqs.~(\ref{densityfluct})-(\ref{phasefluct}). In regime (i), the fluctuations reduce to 
\begin{eqnarray}
\fl \delta \rho_A(x, t)  =   \tilde{\mu} \left[ 1 - \mathrm{tanh}(\tilde{x}) \right]^2 e^{  -  \tilde{\mu} \{ | \tilde{x}|  + \mathrm{sech}^2(\tilde{x}) [ \mathrm{tanh}(\tilde{x}) -1 ] \} } \,    (w-z) \, \mathrm{cos} (Et/\hbar)\, ,  \\
\fl \delta \rho_B(x, t) =   \tilde{\mu} \left[ 1 +  \mathrm{tanh}(\tilde{x}) \right]^2 e^{ -  \tilde{\mu} \{  |\tilde{x}| + \mathrm{sech}^2(\tilde{x}) [ \mathrm{tanh}(\tilde{x}) +1 ]  \} } \,   (q-s)\,  \mathrm{cos} (Et/\hbar) \, ,  
\end{eqnarray}
and
\begin{eqnarray}
  \delta \varphi_A(x, t) = \mathrm{tan}^{-1}\!  \left[ e^{  -  \tilde{\mu} \{  |\tilde{x}| + \mathrm{sech}^2(\tilde{x}) [ \mathrm{tanh}(\tilde{x}) -1 ]  \} } \,   (w + z) \, \mathrm{sin}(E t/\hbar)       \right] \,  , \\
     \delta \varphi_B(x, t)  = \mathrm{tan}^{-1} \! \left[ e^{  -  \tilde{\mu} \{  |\tilde{x}| + \mathrm{sech}^2(\tilde{x}) [ \mathrm{tanh}(\tilde{x}) +1 ] \} }\, (q + s) \, \mathrm{sin}(E t/\hbar)       \right] \, . 
\end{eqnarray}
A bound state zero mode ($E =0$) is supported associated with the breaking of translational symmetry by the soliton~\cite{Lewenstein1996,Castin1998}. In this case, we find that $Q(\tilde{x}) = R(\tilde{x})$ and $S(\tilde{x}) = T(\tilde{x})$ over the infinite interval $- \infty < \tilde{x} < + \infty$ so that $w=z$ and $s=q$. Thus, in the standard normalization the zero mode has a zero norm, $ \int \!dx \, [ | u_{A(B)}|^2 - |v_{A(B)}|^2 ]= 0$, and must instead be normalized according to $\int \! dx\, |u_{A(B)}|^2 = \int \! dx\, |v_{A(B)}|^2$\cite{Dziarmaga2004}. Moreover, very near the core where $|x|/\xi \ll 1$, where $\xi = \hbar c_l/U$ is the healing length which gives the approximate width of the core, the potential functions satisfy the relations $Q \approx T + \mathcal{O}(x/\xi)$, $R \approx  S + \mathcal{O}( x/\xi)$, so that $w \approx s$, $z \approx q$. In this case, the total combined density and phase fluctuations from both sublattices are
\begin{eqnarray}
         \delta \rho(x , t)  =    \delta \rho_{A}(x , t) +  \delta \rho_{B}(x , t)  \approx  0  \,  , \\
         \delta \varphi(x , t)  =   \delta \varphi_{A}(x , t) +  \delta \varphi_{B}(x , t) \nonumber \\
        \hspace{3.5pc}   \approx   2  \,  \mathrm{tan}^{-1}\!  \left[  \mathrm{sin}(E t/\hbar)   + \mathcal{O}(x/\xi)    \right] \, . 
\end{eqnarray}
Thus, near the soliton core the quasi-particle mode contributes to an overall phase fluctuation but does not contribute to density fluctuations for which the A and B sublattice contributions cancel exactly. This is the Nambu-Goldstone mode associated with simultaneous U(1) symmetry breaking in both A and B sublattices. In regime (ii) the situation is slightly different since here the functions $w, \, z, \, q, \, s$ oscillate, but the general results of our analysis hold. Physically, fluctuations of the soliton core correspond to an additional quantum uncertainty in a single measurement of value of the phase at the core and to a nonzero average phase over large time scales which imparts a net translational motion to the soliton.

\section{Continuous spectrum far from the soliton core}
\label{continuousspectrum}

\begin{figure}[t]
\centering
\subfigure{
\label{fig:ex3-a}
\includegraphics[width=.65\textwidth]{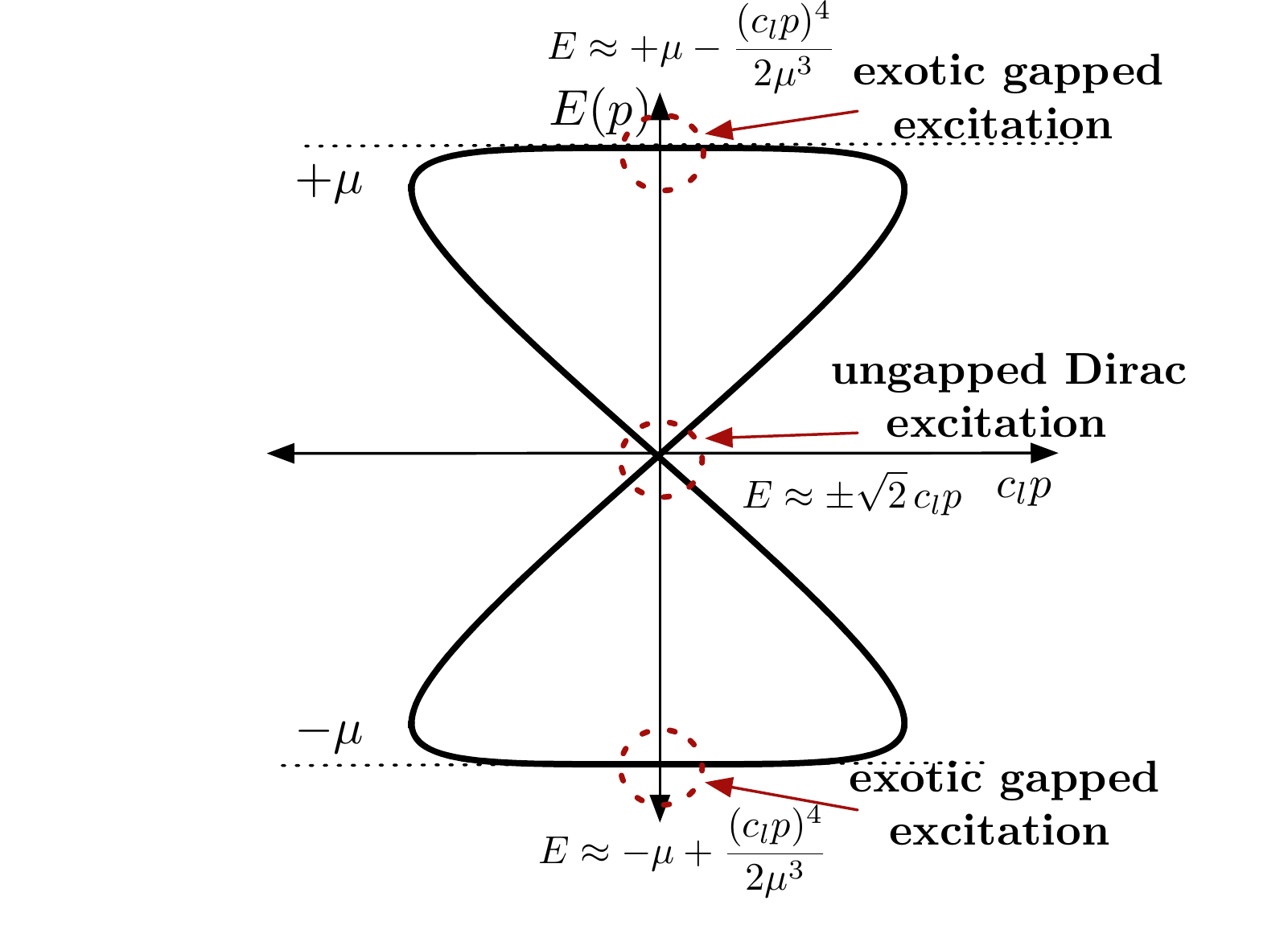}} \\
\caption[]{\emph{Continuous spectrum far from the core of the dark and bright solitons}. The spectrum in this regime consists of linear massless Dirac-like excitations and two types of gapped excitations with (exotic) quartic momentum dependence.  }
\label{ContinuousSpectrum}
\end{figure}

Equations~(\ref{finalform1})-(\ref{finalform2}) and (\ref{finalform3})-(\ref{finalform4}) are useful for obtaining bound states at the soliton core. However, far from the core the backgrounds $\psi_{A(B)}$ are constant and RLSE solutions have plane-wave form. We illustrate this point by solving Eqs.~(\ref{eqn:RLSE1})-(\ref{eqn:RLSE4}) for the zigzag geometry far from the core where $\left|x\right| /\xi \gg 1$. With the quasi-particle state vector written as ${\bf b}(x) =  e^{-i p x/\hbar} \, \left( u_{pA}, \, u_{p B}, \, v_{pA}, \, v_{pB} \right)^T$, the eigenvalue condition becomes
\begin{eqnarray}
  \mathrm{det} \! \left( \begin{array}{ c c c c}
    i c_l  p             & (E -  \mu  )   &  0 &  \mu   \\
        - (   E +    \mu   )           &  i c_l     p   & 0  & 0  \\
                 0     & -   \mu   & i c_l    p    & ( E   +     \mu  )  \\
                   0       &  0  & -(    E   -     \mu   )   &  i c_l    p    \\
                          \end{array} \right)  = 0 \, , 
\end{eqnarray}
which yields the spectrum
\begin{eqnarray}
   E^2(   p ) = \frac{1}{2}    \mu^2 + (c_l     p )^2  \pm \frac{   \mu^2}{2} \sqrt{ 1- \frac{ 4 (c_l    p  )^2}{ \mu ^2}} \, ,  \label{contspec}
\end{eqnarray}
The long wavelength limit of Eq.~(\ref{contspec}) defined by $c_l  p /\mu  \ll 1$, gives the four branches of the continuous spectrum
\begin{eqnarray}
E_g^{\, \pm}(p)  =   \pm \left( \mu - \frac{c_l^4 p^4}{2 \mu^3} + \dots \right)  \, ,  \label{Eg} \\
E_0^{\, \pm}(p)  =   \pm  \sqrt{2} \left(  c_l p +   \frac{ c_l^3 p^3}{4 \mu^2} + \dots \right) \label{E0}  \, , 
\end{eqnarray}
where the two modes $E_g^{\, \pm}$ have a gap equal to the condensate chemical potential $\mu$, and the modes $E_0^{\, \pm}$ are gapless linear Dirac-like excitations for $p \to 0$. The continuous spectrum Eq.~(\ref{contspec}) is plotted in Fig.~\ref{ContinuousSpectrum}. We point out that the results in this section apply to the dark and bright solitons alike since both share the same asymptotic form.

\section{Relativistic solitons as domain walls}
\label{Pseudospin}

In the case of NLDE solitons the asymptotic values of the spin components correspond to two different asymptotically flat vacua connected through the core region where a spin twist occurs. In this section we study the internal spin space rotations that occur inside the core regions of NLDE solitons, and thus firmly establish such solutions as domain walls. We work in the chiral representation since this is the most natural approach as particle interactions do not mix upper and lower Dirac two-spinors. For the two-spinor order parameter $\psi(x, t) = \left[ \psi_A(x, t), \, \psi_B(x, t) \right]^T$, the quasi-1D reduction of the NLDE is expressed concisely in terms of spin-orbit coupling and contact interaction terms
\begin{eqnarray}
\left( i \hbar \sigma^\mu  \partial_\mu  + \mathrm{N}  \right) \psi  = 0  \, ,  \label{spinorNLDE}
\end{eqnarray}
 where the interactions are contained in the $2 \times 2$ diagonal matrix $\mathrm{N}$ defined as
\begin{eqnarray}
\mathrm{N}  \equiv \left( \begin{array}{ c c} 
   U  |\psi_A|^2  &  0  \\
     0 &U  |\psi_B|^2  \\
     \end{array}  \right)  \, ,   \label{interactionmatrix}
\end{eqnarray}
and the commonly used ``spin-orbit'' terminology refers to the derivative terms expressed as a contraction between the Pauli matrices $\sigma^\mu$ and the space-time derivatives $\partial_\mu$. The contraction is over the Greek index $\mu = 0, 1$ following the usual Einstein convention $\sigma^\mu \partial_\mu = g^{\mu \nu } \sigma_\mu \partial_\nu$, with the $(1+1)$-dimensional Minkowski metric and Pauli matrices 
\begin{eqnarray}
g^{\mu \nu } =   \left( \begin{array}{ c c } 
  1 &  0   \\
     0   &  -1  \\
     \end{array}  \right) , \;  \; \sigma^{0} =   \left( \begin{array}{ c c } 
  1 &  0   \\
     0   &  1  \\
     \end{array}  \right) \, . 
     \end{eqnarray}
The additional Pauli matrix $\sigma^1$ is chosen for the specific case of the real or complex Dirac operator, 
  \begin{eqnarray}
      \sigma^{1}_{\mathrm{real}}=   \left( \begin{array}{ c c } 
  0 &  -i    \\
    i   & 0  \\
     \end{array}  \right) , \; \;       \sigma^{1}_{\mathrm{complex}}=   \left( \begin{array}{ c c } 
  0 &  1   \\
    1  & 0  \\
     \end{array}  \right) \, . 
\end{eqnarray}
A thorough description of this formalism may be found in Ref.~\cite{haddad2009}.

To make the connection between superfluidity and relativistic current apparent in the pseudospin formalism, we use the Gordon decomposition approach~\cite{bjorken64} which lets us separate the phase gradient from the magnitude gradient in Eq.~(\ref{spinorNLDE}). The former describes conventional superfluidity and the latter contains the pseudospin dependent contribution to the overall current. In our previous work~\cite{haddad2009} we found that the NLDE has the associated current
\begin{eqnarray}
j^\mu =   \bar{\psi} \, \sigma^\mu\,  \psi \, ,   \label{NLDEcurrent}
\end{eqnarray}
where $\bar{\psi} \equiv \psi^\dagger \sigma^0$. Solving Eq.~(\ref{spinorNLDE}) for $\psi$ and the conjugate of Eq.~(\ref{spinorNLDE}) for $\bar{\psi}$ allows us to express Eq.~(\ref{NLDEcurrent}) as 
\begin{eqnarray}
 j^\mu = \frac{1}{2}\left[     \bar{\psi} \, \sigma^\mu \mathrm{N}^{-1}  \sigma^\nu       \left( \partial_\nu  \psi \right)  -    \left(   \partial_\nu  \bar{\psi}   \right)  \sigma^\nu \mathrm{N}^{-1}  \sigma^\mu \psi   \right] \, ,  \label{NLDEcurrent2}
\end{eqnarray}
where $\mathrm{N}^{-1}$ is the inverse matrix of Eq.~(\ref{interactionmatrix}). Working out the properties of the matrix products $\sigma^\mu \mathrm{N}^{-1}  \sigma^\nu$, we find the decomposition 
\begin{eqnarray}
\sigma^{\mu \nu } = g_\alpha^\mu \, D^{\alpha \nu}  + O^{\mu \nu} \, 
\end{eqnarray}
where $D^{\alpha \nu}$ and $ O^{\mu \nu}$ are $2 \times 2$ diagonal and off-diagonal matrices with respect to the spin indices, respectively, and encapsulate the interaction terms. The current in Eq.~(\ref{NLDEcurrent2}) can now be expressed as
\begin{eqnarray}
j^\mu &=   \frac{1}{2}\left(     \bar{\psi} \, g_{\alpha}^\mu  D^{\alpha \nu}  \,    \partial_\nu^{\pm} \,    \psi  +     \bar{\psi} \, O^{\mu \nu}  \,     \partial_\nu^\pm \,    \psi      \right)  \nonumber  \\
   &=  j^\mu_\mathrm{orb} \, +\,  j^\mu_\mathrm{pspin}   \;  ,  \label{NLDEcurrent3}
\end{eqnarray}
where we use the $\partial_\nu^\pm$ notation to mean the antisymmetric differentiation $\bar{\psi} \,M^{\mu \nu}  \partial_\nu^\pm   \psi  \equiv  \left( \partial_\nu \bar{\psi} \right)  M^{\mu \nu}  \psi - \bar{\psi} M^{\mu \nu} \left(  \partial^\mu \psi \right)$. The first term in Eq.~(\ref{NLDEcurrent3}) is the orbital contribution to the current which does not mix pseuodspin indices and describes the familiar superfluid velocity as the gradient of an overall phase, whereas the second term is the pseudospin portion which mixes spinor components and describes the phase-independent density current.

For solutions of the quasi-1D NLDE, Eq.~(\ref{NLDEcurrent3}) gives the following results: 

\begin{enumerate}[label=(\roman{*})]

\item For the real Dirac operator with the spinor solution $\Psi(x, t) = e^{i \mu t/\hbar}  \left[ i  f_A(x), \, f_B(x) \right]^T$, where $f_A$ and $f_B$ are real functions of $x$ and  solve the time-independent armchair NLDE, we have 
\begin{eqnarray}
j^\mu_\mathrm{orb}  &=   0 \, , \\
j^t_\mathrm{pspin}   &=   f_A^2 + f_B^2 \, , \\
j^x_\mathrm{pspin}  &=   - 2 f_A f_B \, . 
\end{eqnarray}

\item For the complex Dirac operator the relative phase between spinor components is zero and $\Psi(x, t) = e^{i \mu t/\hbar}  \left[  f_A(x), \, f_B(x) \right]^T$ for which we find
\begin{eqnarray}
j^\mu_\mathrm{orb}  &=  0 \, , \\
j^t_\mathrm{pspin}  &=  f_A^2 + f_B^2 \, ,  \\
j^x_\mathrm{pspin}  &=  2 f_A f_B \, . 
\end{eqnarray}

\end{enumerate}
Thus, for the ordinary dark soliton and bright solitons found in Ref.~\cite{haddadcarrsoliton1}, the currents are peaked at the soliton cores with the current $j^x_\mathrm{pspin}$ for solitons associated with the real Dirac operator pointing in the negative $x$ direction and in the positive direction for the complex case. To understand results (i) and (ii) we emphasize that NLDE solitons have a constant (spatially uniform) overall phase which reflects the zero superfluid current $j^\mu_\mathrm{orb} =0$, whereas the relative phase between spinor components varies through the soliton core leading to nonzero values for $j^t_\mathrm{pspin}$ and $j^x_\mathrm{pspin}$.

The structure of NLDE solitons is reminiscent of domain walls as in the interface between $^3\mathrm{He}$-A and $^3\mathrm{He}$-B superconductors~\cite{Vollhardt2002} but more closely resembling the boundary between pseudospin domains in two-component BECs. The analogy extends only loosely to spin-1 BECs~\cite{Watabe2012} in the case where the asymptotic order parameter (far from the domain wall, i.e., $x \to \pm \infty$) is in oppositely polarized states. However, we must point out that two-dimensional BECs in honeycomb lattices are Dirac spin-1/2 analogs which couple spinor components through the kinetic term; in contrast, multi-component BECs are coupled through the interaction term.

Our system possesses a pseudospin structure described by the vector operator ${\bf S} = ( S_x,  S_y,  S_z)$, where the component spin operators are expressed in terms of the Pauli matrices $S_{x, y, z} =  (\hbar/ 2 )\,  \sigma_{x, y, z}$, and the spin quantization axis is in the z-direction. The appearance of the Pauli matrices here underscores the marked contrast to ordinary two-component BECs: the latter are not associated with a Clifford algebra, i.e., fermionic anticommutativity. Although strictly speaking there is no z-component of spin in our problem and we must choose between $\sigma_x$ or $\sigma_y$ for the two types of quasi-1D NLDEs (real or complex Dirac operator), the transverse spin operator $\sigma_z$ is still relevant in characterizing our solutions.

The measurable physical quantity is the average pseudospin or spin density defined as $\hat{\bf F}(x, t)= \hat{\psi}^\dagger(x, t) {\bf S} \hat{\psi}(x, t)$. In the case where the BEC occupies the A sublattice only, i.e., $\Psi(x) = [ \psi_A(x),\,  0]^T$, we have
\begin{eqnarray}
{\bf F}(x) &=\Psi^*(x) {\bf S} \Psi(x)  \\
             &=    \Psi^*(x) \left(  S_x  , \, S_y  , \,   S_z  \right)  \Psi(x)  \\
             &= ( +  \hbar/2 )  |\psi_A(x)|^2  \left( 0, \, 0 , \, 1 \right) \, , 
\end{eqnarray}
whereas occupation of the B sublattice has ${\bf F}(x)= (-  \hbar/2 ) |\Psi_B(x)|^2  \left( 0, \, 0 , \, 1 \right) $. The total particle density of our solitons varies through the core (the dark soliton dips, the bright soliton peaks), but returns to the constant value $\lim_{ x/ \xi \to \pm \infty}\rho  = + \mu/U$ with $\mu/U =1$ for NLDE solitons. In contrast, the spin density has asymptotic values $\lim_{x/\xi \to- \infty} {\bf F} =  ( \hbar / 2) ( 0, \, 0,\, 1)^T$, $\lim_{x/\xi \to \infty} {\bf F} =  ( \hbar  / 2) ( 0, \, 0,\, -1)^T$. This allows for the interpretation of NLDE solitons as boundaries or domain walls separating regions of spin +1/2 and spin -1/2, with the transition between the two asymptotic regions taking place within the soliton core.

\section{Spin waves and Nambu-Goldstone modes}
\label{QuadSpinWaves}

In this section we will extend the discussion from Sec.~\ref{Pseudospin} to the asymptotic linear spectrum far from the soliton core. As discussed in Eqs.~(\ref{psisplit1})-(\ref{psisplit3}), in general the quantum perturbed wave function can be written as $\Psi(x) + \hat{\phi}(x , t) = [\psi_A(x) +  \hat{\phi}_A(x, t), \, \psi_B(x) + \hat{\phi}_B(x, t) ]^T$, from which the fluctuation in the spin density $\delta {\bf F} =( \delta F_x, \,  \delta F_y, \, \delta F_z )^T$ is computed to be
\begin{eqnarray}
\delta F_x  &\equiv (\hbar/2) (\psi_A^* \phi_B + \phi_A^* \psi_B  + \psi_A \phi_B^* + \phi_A  \psi_B^*)\,  ,  \label{Fx} \\
\delta F_y   &\equiv - i (\hbar/2) (\psi_A^* \phi_B + \phi_A^* \psi_B  -  \psi_A \phi_B^* -  \phi_A  \psi_B^*) \, ,  \label{Fy} \\
\delta F_z  &\equiv (\hbar/2)  (\psi_A^* \phi_A + \phi_A^* \psi_A  -  \psi_B^* \phi_B -  \phi_B^*  \psi_B) \, . \label{Fz} 
\end{eqnarray}
Note that we have used the quasiparticle averages $ \phi_{A(B)} = \langle \hat{\phi}_{A(B)} \rangle$ and condensed the notation by not writing out the space-time dependence of the component fluctuations. In the case of a spin-1 BEC there are three types of spin fluctuations: transverse, quadrupolar, and Nambu-Goldstone. The first mixes the $S_z = \pm 1$ into the $S_z= 0$ hyperfine channel, the second mixes the $S_z = +1$ and $-1$ hyperfine states, and the third is the phase fluctuation associated with $U(1)$ symmetry breaking (see Ref.~\cite{Watabe2012}).

 In our spin-1/2 problem, the physical picture is slightly different. Here there are only two modes: a spin wave mode which mixes spin-up and spin-down states and the Nambu-Goldstone mode. For the real and complex Dirac operator, the spin wave modes are, respectively, $\delta F_\mathrm{SW} =  \delta F_x$ and $\delta F_\mathrm{SW} = \delta F_y$; in both cases the Nambu-Goldstone mode is $\delta F_{\mathrm{NG}} = \delta F_z$. Near the soliton core, the spin wave mode vanishes as we showed in Sec.~\ref{BoundFluctuations}, leaving only the Nambu-Goldstone mode. Far from the core the situation is different. For instance, for the real Dirac operator where $x/\xi \to - \infty$, we have $F_z = +\hbar/2$, $\delta F_\mathrm{SW} =  (\hbar/2) ( \phi_B  +  \phi_B^*)$, and $\delta F_{\mathrm{NG}} =  (\hbar/2)  ( \phi_A + \phi_A^* )$. In contrast, in the other limit $x/\xi \to + \infty$, $F_z = -\hbar/2$, $\delta F_\mathrm{SW} =  (\hbar/2) ( \phi_A  +  \phi_A^*)$, and $\delta F_{\mathrm{NG}} =  (\hbar/2)  ( \phi_B + \phi_B^* )$.

 We may verify that our definitions for $\delta F_\mathrm{SW}$ and $\delta F_{\mathrm{NG}}$ make sense physically by computing their momentum dependence. Expanding the quasi-particle functions gives $\phi_{A(B)}(x, t; p) = \exp \{- i [ px + E(p)t]/\hbar \} [ u_{A(B)}(p) - v_{A(B)}^*(p) ] $, and solving the RLSE in the long wavelength limit ($c_lp/\mu \ll 1$), shows that the gapless excitation, with energy $E_0^{\pm}$ described by Eq.~(\ref{E0}), corresponds to $u_{A}(p) = v_{A}(p) =0$ and $u_{B}(p) = v_{B}(p)$, so that $\delta F_\mathrm{SW} =  0$ and $\delta F_{\mathrm{NG}} =   \hbar \, \mathrm{Im}[ u_B(p)]$. Following similar steps we find that the gapped excitations $E_g^{\pm}$ in Eq.~(\ref{Eg}) give $\delta F_\mathrm{SW}^+ =  \hbar\,  \mathrm{Re}[v_A(p)]$ and $\delta F_{\mathrm{NG}}^+ =  0$, and similarly $\delta F_\mathrm{SW}^- =  \hbar \, \mathrm{Re}[u_A(p)]$ and $\delta F_{\mathrm{NG}}^- =  0$. Thus, $F_\mathrm{SW}$ and $F_\mathrm{NG}$ do indeed correspond respectively to the gapped and gapless modes  of the $S = - 1/2$ and $S= +1/2$ domains.

\begin{figure}[t]
\centering
\subfigure{
\label{fig:ex3-a}
\includegraphics[width=.65\textwidth]{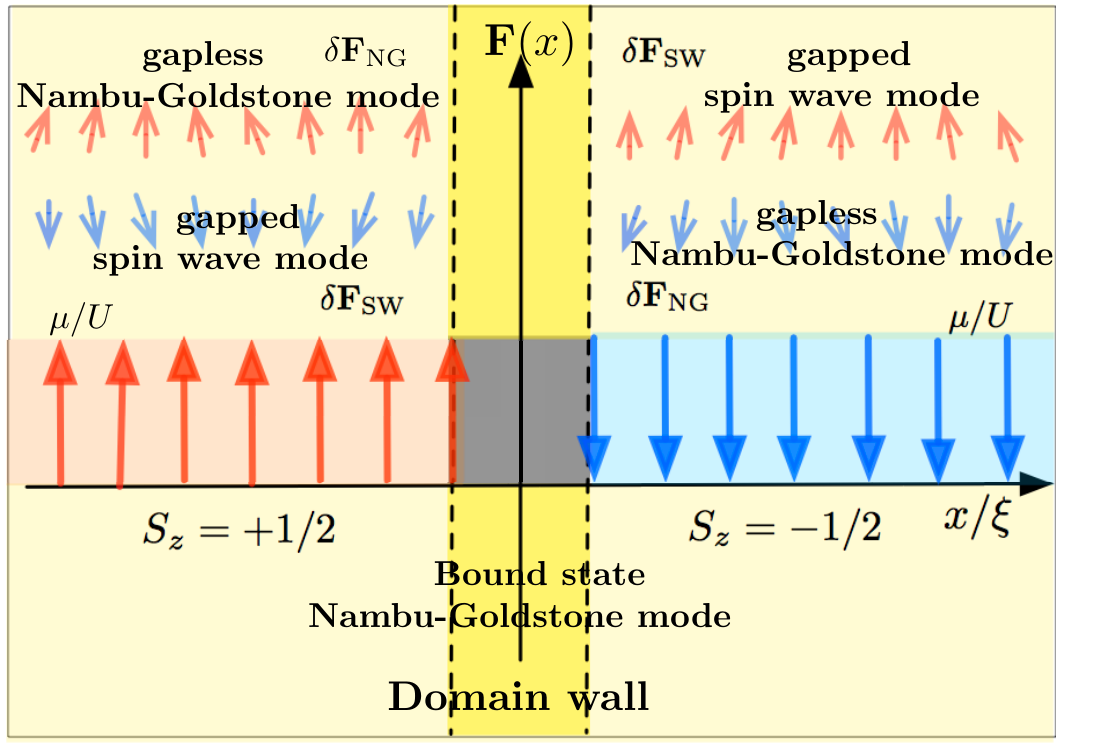}} \\
\caption[]{\emph{Sketch of pseudospin domain structure of NLDE solitons}. The central region (bright yellow) is the soliton core and divides two asymptotic spin domains (blue and red). Inside the core, sublattice density fluctuations cancel each other leaving only the Nambu-Goldstone phase fluctuation. On either side of the core, the gapped mode introduces admixtures of the two spin polarizations and corresponds to fluctuations in the condensate density. The Nambu-Goldstone mode introduces admixtures in the phase of each polarization.}
\label{PseudospinDomain}
\end{figure}

The formulation of fluctuations in terms of spin wave modes sets the ground work for detailed analysis of scattering from the soliton core, an interesting topic in itself which may provide insight into the question of integrability of the quasi-1D reduction of the NLDE. The pseudospin domain structure which we have discussed in this section is summarized in Fig.~\ref{PseudospinDomain}.

\section{Discrete spectra for solitons in a harmonic trap}
\label{SolitonSpectra}

In experiments, solitons reside in a BEC within a harmonic magnetic trap. Consequently, in the case of extended  solitons the trap boundary affects the soliton in a nontrivial way. The result is quantization of spatial modes, which has a significant effect at large healing length (comparable to the trap size) or equivalently for weak nonlinearity. In this section we study the behavior of our dark and bright solitons in the presence of a harmonic trap by computing the chemical potential spectra for the single, double, and triple soliton states. 

% As throughout this Article, these results apply to both the zigzag and armchair geometries. 

 For the case of a highly oblate harmonic confining potential which defines a 2D system, the oscillator frequencies satisfy $\omega_z \gg \omega \equiv  \omega_x , \, \omega_y$. If in addition to this condition we also take $\omega_y \gg \omega_x$, as we discussed in \cite{haddadcarrsoliton1} with the soliton in the $x$-direction, we encounter only the effects resulting from quantization of spatial modes along the soliton direction shown in Fig.~\ref{OneD}. We then take the longitudinal trapping potential to be $V(x) = \frac{1}{2} M\, \omega^2 x^2$. We proceed numerically by incorporating this potential into the NLDE and then transforming to dimensionless equations by defining
\begin{eqnarray}
\chi \equiv \frac{ \hbar \omega \, x }{\hbar c_l}\; , \;\;\; \eta_{A(B)} \equiv \sqrt{ \frac{U}{\hbar \omega}} \, f_{A(B)}\, , \label{chi}
\end{eqnarray}
thereby obtaining the dimensionless form of the NLDE
\begin{eqnarray}
- \partial_{\chi}  \eta_B(\chi) +  \left|\eta_A(\chi)\right|^2\! \eta_A(\chi)  + \mathcal{Q} \,  \chi^2 \, \eta_A(\chi)  &= \tilde{\mu} \,  \eta_A(\chi) \, ,  \label{dimensionless3}    \\
    \partial_{\chi} \eta_A(\chi) +  \left| \eta_B(\chi)\right|^2 \! \eta_B(\chi)    + \mathcal{Q} \,  \chi^2 \,  \eta_B(\chi) &=   \tilde{\mu} \,  \eta_B(\chi) \, . \label{dimensionless5}
\end{eqnarray}
Here the two rescaled physical parameters in the NLDE are
\begin{eqnarray}
 \mathcal{Q} \equiv  \frac{M c_l^2}{2\, \hbar \omega} \; , \;\;\;\; \;\;\;\;\;  \tilde{\mu} \equiv \frac{\mu}{\hbar \omega} \, . \label{Qdef}
\end{eqnarray}
The analogous rescaling for the case of the NLSE in an oblate harmonic trap differs from our problem in a fundamental way. For the NLSE, energies are scaled to the trap energy $\hbar \omega$ and lengths to the oscillator length $\ell = \sqrt{\hbar/M \omega}$ (see Ref.~\cite{CarCla06}). In contrast, our problem retains the same scaling to the trap energy but lengths are scaled to the ratio $c_l/\omega = t_h a \sqrt{3}/2\hbar \omega$ as can be seen in Eq.~(\ref{chi}), where the natural scales of the lattice appear in the hopping energy $t_h$ and lattice constant $a$. This particular choice of scaling is forced on us because of the single spatial derivative in the NLDE; we cannot completely scale away the lattice information. The mass energy factor $M c_l^2$ in Eq.~(\ref{Qdef}) is thus a direct result of the relativistic linear dispersion of the NLDE. Length scales are defined in Table~\ref{scaletable} along with associated momentum and energy scales for quasi-1D NLDE solitons in a harmonic trap. Note that the last three scales in Table~\ref{scaletable} are key in our calculations since they contain information about the first two scales. 
\begin{table}[h]
 \begin{indented}
\item[]\begin{tabular}{@{\hspace{.75pc}}llll}
\br
\lineup
 Physical scale  &  Length   &  Momentum & Energy   \\
 \mr
  nonlinear &    $\xi = 1 /\sqrt{8 \pi \bar{n} a_s}$  &  $\sqrt{ 8 \pi \hbar^2 \bar{n} a_s }$&   $8 \pi \hbar^2 \bar{n} a_s /M$    \\
  
  transverse     &    $\ell_\perp= \sqrt{\hbar /M \omega_\perp}$  & $\sqrt{ \hbar  M \omega_\perp}$   & $\hbar \omega_\perp$    \\

  chemical potential     &    $\ell_\mu =  \hbar c_l/\mu$  & $\mu / c_l$   & $\mu^2/M c_l^2$    \\
 
  lattice   &    $\ell_\mathrm{latt}= \hbar /Mc_l$  &  $M c_l$& $Mc_l^2$    \\

   quasi-1D     &  $\xi_\mathrm{1D} = \hbar c_l /U_\mathrm{1D}$ &  $U_\mathrm{1D} /c_l$    &    $U_\mathrm{1D}^2/M c_l^2$         \\

  harmonic trap    &  $\ell_\mathrm{trap} = \sqrt{\hbar /M\omega} $  & $\sqrt{ \hbar  M \omega}$  &  $\hbar \omega$         \\

 \br 
\end{tabular}   
{\caption{\emph{Physical Scales}. Length, momentum, and energy scales for the quasi-1D NLDE in a harmonic trap. Scales are determined by the 3D healing length $\xi$; the transverse oscillator length $\ell_\perp$; the large-momentum healing length $\ell_\mu$; the scale $\ell_\mathrm{latt}$ associated with the lattice constant and hopping energy; the low-momentum quasi-1D healing length $\xi_\mathrm{1D}$; and the harmonic trap length $\ell_\mathrm{trap}$. All other fundamental parameters were defined in Sec.~\ref{SolitonStability}. Momentum and energy scales are related to their associated length scales by: momentum $\sim \hbar / \mathrm{length}$ and energy $\sim \hbar^2 / (M \times \mathrm{length}^2)$. We have included the transverse oscillator frequency $\omega_\perp$ which defines the transverse size of the condensate, either in the direction normal to or along the width of the nanoribbon .    }\label{scaletable}}
 \end{indented}
\end{table}
\noindent Realization of the NLDE solitons in a harmonic trap requires the particular ordering of length scales 
\begin{eqnarray}
\ell_\mathrm{latt} \ll \xi_\mathrm{1D} \ll \ell_\mathrm{trap} \, ,   \label{scales}
\end{eqnarray}
due to the long-wavelength approximation used to obtain the NLDE.$^{\footnotemark[1]}$ The lengths in the hierarchy Eq.~(\ref{scales}) are: the lattice scale $\ell_\mathrm{latt}$, which contains the lattice constant and hopping energy; the quasi-1D effective healing length $\xi_\mathrm{1D}$, which incorporates the atom-atom interaction and the transverse length; and the harmonic trap length $\ell_\mathrm{trap}$, which defines the overall size of the BEC. For a typical scenario for a $^{87}$Rb BEC~\cite{haddadcarrsoliton1} with a trap oscillator frequency $\omega = \mathrm{2}\pi \times 0.039 \, \mathrm{Hz}$, we obtain $\ell_\mathrm{latt} \approx 2.3 \,   \mathrm{ \mu m}$, $ \xi_\mathrm{1D} \approx 10 \,  \mathrm{ \mu m}$, and $\ell_\mathrm{trap} \approx 55 \,  \mathrm{ \mu m}$. 

To connect with dark and bright soliton solutions of the NLDE~\cite{haddadcarrsoliton1} we consider the limit for zero trap energy. This amounts to taking the trap size to infinity, i.e, letting $\ell_\mathrm{trap}$ to be the largest length scale in Table~\ref{scaletable}. Expressed in terms of the lengths in Table~\ref{scaletable} one finds that the derivative, interaction, and chemical potential terms in Eqs.~(\ref{dimensionless3})-(\ref{dimensionless5}) scale as $\ell_\mathrm{trap}/\ell_\mathrm{latt}$, $\ell_\mathrm{trap}^2/\ell_\mathrm{latt} \xi_\mathrm{1D}$, and $\ell_\mathrm{trap}^2/\ell_\mu \ell_\mathrm{latt}$, respectively, while the harmonic term does not scale with the trap size. Thus in the large trap limit the harmonic term can be neglected and we regain the continuum theory as expected.

\begin{figure}[t]
\centering
 \subfigure{
\label{fig:ex3-a}
 \includegraphics[width=.65\textwidth]{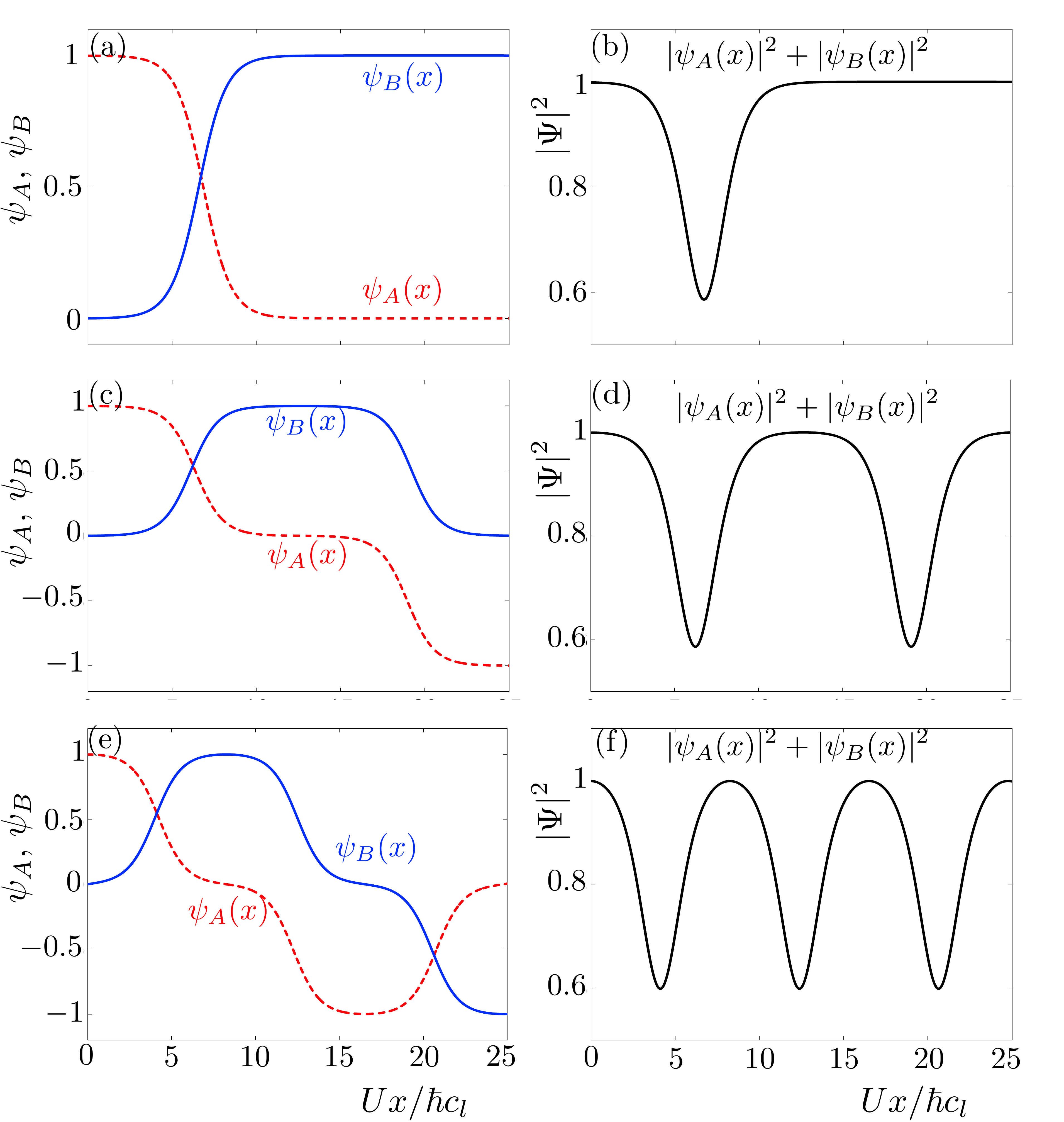}}\\ 
\caption[]{\emph{Multiple dark solitons in the limit of a very weak trap}. (a,b) Single dark soliton. (c,d) Double dark soliton. (e,f) Triple dark soliton. Panels on the left show the A (red) and B (blue) sublattice excitations obtained using a numerical shooting method. The corresponding densities are shown in the right panels. The plots here correspond to the case of the real Dirac operator in the quasi-1D NLDE, with spinor components interchanged for the complex case and no change in the density.} \label{LineSoliton}
\end{figure}

\begin{figure}[t]
\centering
 \subfigure{
\label{fig:ex3-a}
\includegraphics[width=.65\textwidth]{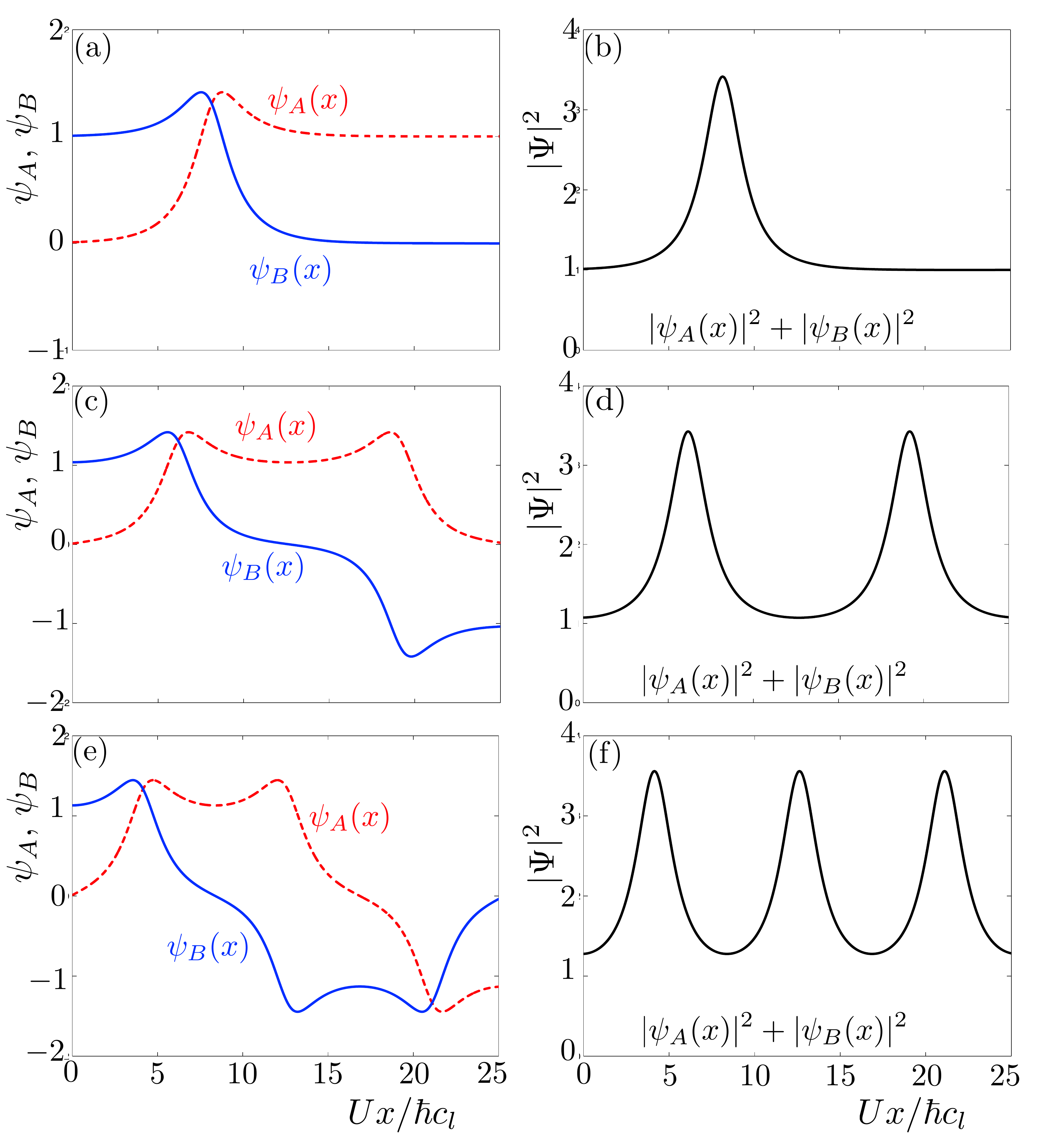}}\\ 
\caption[]{\emph{Multiple bright solitons in the limit of a very weak trap}. (a,b) Single bright soliton. (c,d) Double bright soliton. (e,f) Triple bright soliton. Panels on the left show the A (red) and B (blue) sublattice excitations obtained using a numerical shooting method. The corresponding densities are shown in the right panels. Spinor components shown are for the case of the real Dirac operator in the quasi-1D NLDE and are interchanged for the complex case.} \label{BSolitons}
\end{figure}

We use a numerical shooting method to solve the NLDE in the presence of the harmonic trap~\cite{haddadcarrsoliton1}. This is done by first expanding the spinor wavefunction $\Psi(x) = \left[  \psi_A(x) , \, \psi_B(x)  \right]^T$ in a power series about the center of the trap at $x =0$. The leading coefficient $a_0$ in the expansion for $\psi_A$ is then tuned to obtain a stable solution. The second free parameter $b_0$, from $\psi_B$, is held fixed between 0 and 1, for the dark soliton, or between 1 and the value at the peak, for the bright soliton. For $b_0 = 0$, iterating $a_0$ leads to the soliton solution at $a_0 = a_0^{\mathrm{soliton}}$, where higher precision in $a_0^{\mathrm{soliton}}$ pushes oscillations out to larger values of $x$. Panels (a),(c) and (e) in Fig.~\ref{LineSoliton} show spinor components for single, double, and triple dark solitons with corresponding densities in panels (b), (d) and (f). Analogous plots for the bright soliton are shown in Fig.~\ref{BSolitons}. Nodes only appear in the spinor component functions for the case of multiple solitons but not for single ones, but in every case the total density never drops to zero. The following data was used to obtain the dark solitons in Fig.~\ref{LineSoliton}: $a_0 =  0.9949684287783 \pm 10^{-7}$, $\tilde{\mu} =1$, for the single soliton; $a_0 =  0.99496892372588591202$, $\tilde{\mu} =1.00000103$, for the double soliton; and, $a_0=  0.993 \pm 10^{-17}$, $\tilde{\mu} =1.001$, for the triple soliton. The bright solitons in Fig.~\ref{BSolitons} are associated with the following data: $a_0=  0.010\pm 10^{-17}$, $\tilde{\mu} =1$, for the single soliton; $a_0=  0.11 \pm 10^{-18}$, $\tilde{\mu} =1.04$, for the double soliton; and, $a_0=  0.1 \pm 10^{-19}$, $\tilde{\mu} =1.15$, for the triple soliton. The solutions are converged to the last digit in the numerical values for $a_0$. Greater precision in the value of $a_0$ is required in the case of the single soliton in order to push excitations out to larger values of $x$. See also~\cite{Carr2006} for a study of precision issues in shooting methods related to BEC in harmonic traps.

\footnotetext[1]{Condition (\ref{scales}) can be overcome by turning to a discrete model but still working in the mean field approximation~\cite{haddad2009}.}

\begin{figure}[t]
\centering
\subfigure{
\label{fig:ex3-a}
\includegraphics[width=.65\textwidth]{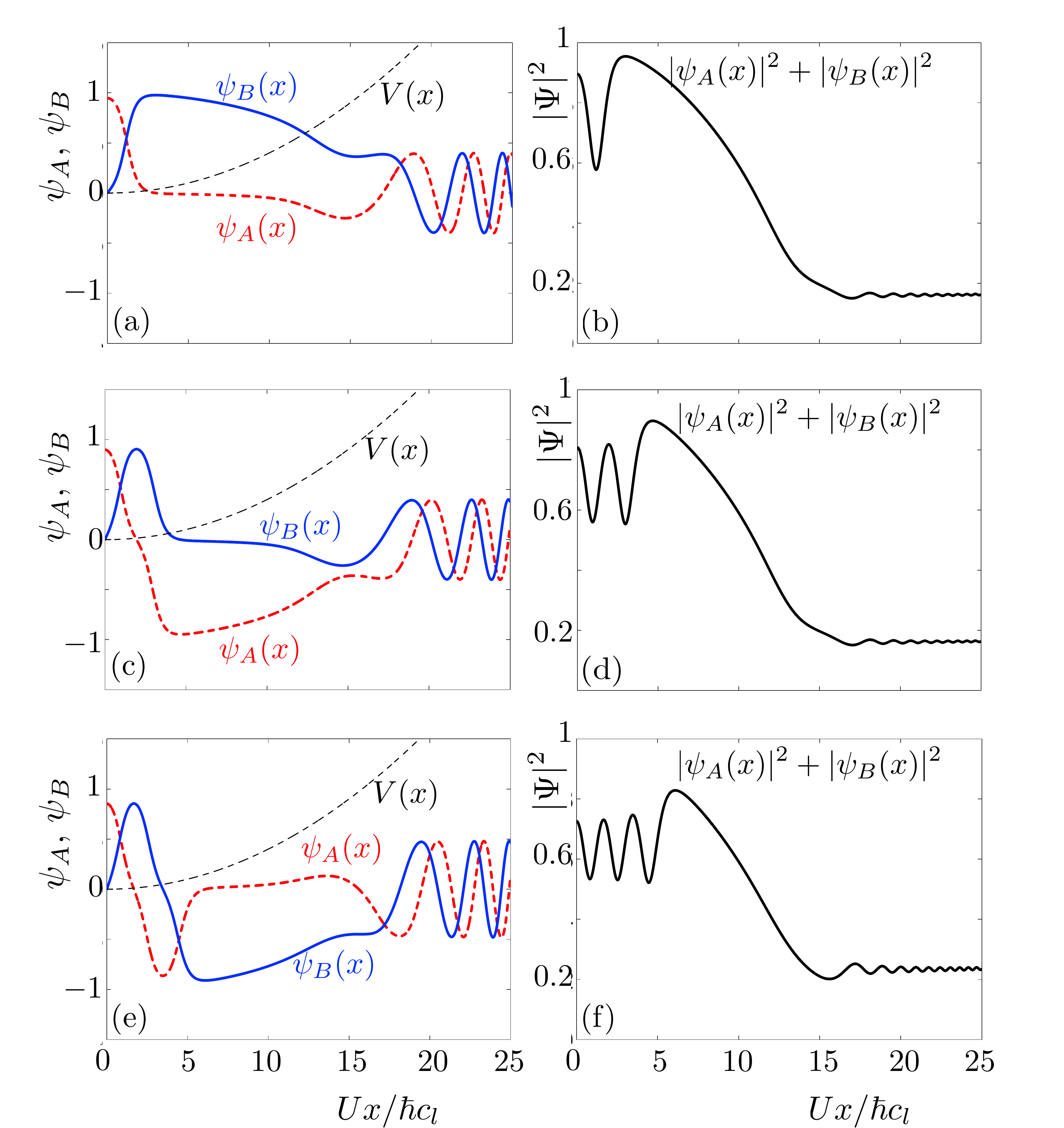}}\\ 
\caption[]{\emph{Multiple dark solitons in a harmonic trap}. (a,b) Single dark soliton. (c,d) Double soliton. (e,f) Triple soliton. Spinor components are shown in left hand panels with corresponding densities shown in the right panels. The black dashed plot is the harmonic trapping potential. } \label{DarkSolitonTrap}
\end{figure}

\begin{figure}[t]
\centering
 \subfigure{
\label{fig:ex3-a}
\includegraphics[width=.65\textwidth]{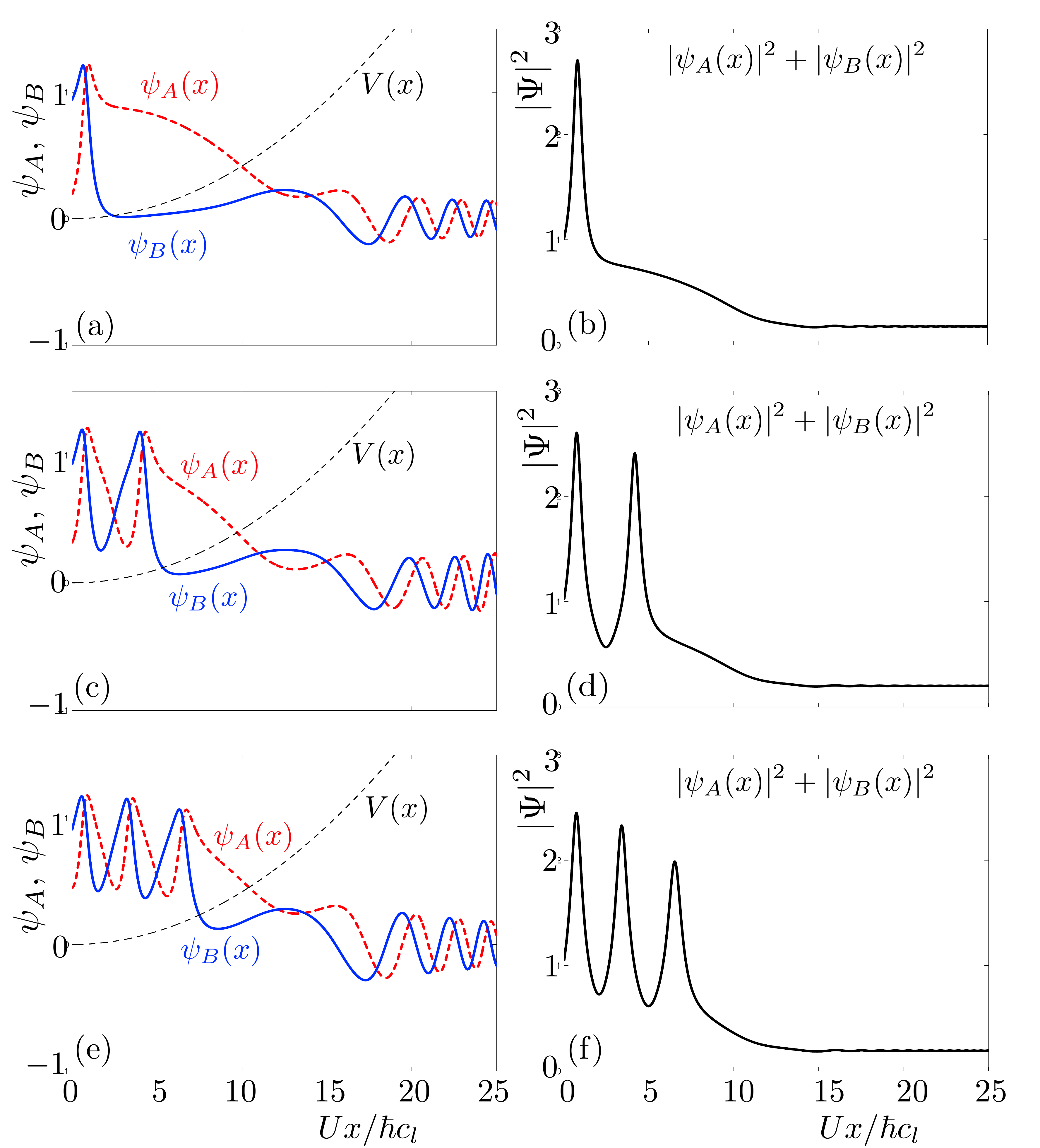}}\\ 
\caption[]{\emph{Multiple bright solitons in a harmonic trap}. (a,b) Single bright soliton. (c,d) Double soliton. (e,f) Triple soliton. Spinor components are shown in left hand panels with corresponding densities shown in the right panels. The black dashed plot is the harmonic trapping potential. } \label{BrightSolitonTrap}
\end{figure}

Next, we solve Eqs.~(\ref{dimensionless3})-(\ref{dimensionless5}) with a nonzero oscillator length. In the presence of the trap potential solutions are spatially quantized and labeled by a discrete index. In particular, for $\mathcal{Q}=10^3$, corresponding to a longitudinal oscillator frequency $\omega = 2 \pi \times  0.0305\,  \mathrm{Hz}$, we find the free parameter $a_0$ for the single dark soliton at $a_0 = 0.94640402384 \pm 10^{-9}$ and for the two and three soliton states $a_0=0.89882708125 \pm 10^{-9}$ and $a_0= 0.8523151\pm 10^{-13}$, respectively. The lowest multiple dark solitons in a trap are plotted in Fig.~\ref{DarkSolitonTrap} along with their corresponding densities. The minima near the origin in the density plots, Figs.~\ref{DarkSolitonTrap}(b), (d), (f), correspond to the density notch in the unconfined case, Figs.~\ref{LineSoliton}(b), (d), and (f). The number of notches identifies the single, double, and triple soliton states. Analogous plots for the bright soliton are displayed in Fig.~\ref{BrightSolitonTrap}.

It is worth commenting on the oscillating behavior in the tails of the spinor components in Figs.~\ref{DarkSolitonTrap}-\ref{BrightSolitonTrap}. Oscillations such as these in large  potential regions are fundamentally inherent to the Dirac equation. To clarify the source of this effect, we rewrite Eqs.~(\ref{dimensionless3})-(\ref{dimensionless5}) as
\begin{eqnarray}
 \eta_B'  =  \left(  \mathcal{Q} \,  \chi^2  - \tilde{\mu} +| \eta_A|^2 \right) \eta_A \, , \label{}                       \\
 \eta_A'  =  - \left(  \mathcal{Q} \,  \chi^2   - \tilde{\mu} +|\eta_B|^2 \right) \eta_B  \label{}   .
\end{eqnarray}
Near the origin, the trap potential is weak and the chemical potential term dominates so that we have $\eta_B' < 0$ and $\eta_A' > 0$. However, as we move away from the origin and into the strong potential region the quadratic terms in $\chi$ grow eventually overwhelming the other terms. In this asymptotic region $\eta_A$ and $\eta_B$ solve the limiting equations
 \begin{eqnarray}
 \eta_B'  =     \mathcal{Q} \,  \chi^2  \,  \eta_A  \,  , \; \;\;\;  \eta_A'    =  -   \mathcal{Q} \,  \chi^2  \,  \eta_B  \label{asymptotic2}  , 
\end{eqnarray}
whose solutions are
 \begin{eqnarray}
 \eta_B(\chi)  =   \frac{1}{3}\,  \mathrm{sin} \left[ (\mathcal{Q} \,  \chi^2) \chi \right] \,   , \;\;\;\;  \eta_A(\chi) =     \frac{1}{3}\, \mathrm{cos} \left[ (\mathcal{Q}   \chi^2) \chi \right]   . \label{asymptotic4}   
\end{eqnarray}
These functions oscillate with a spatially increasing frequency $k \equiv \mathcal{Q} \,\chi^2$, so it is clear that the tail oscillations are coming from the unbounded potential barrier. Physically, the barrier potential forces a positive energy particle into the continuum of negative energy states below the Dirac point. In contrast, this effect does not arise for an ordinary Schr\"odinger-like particle in a quasi-1D harmonic potential. There the particle is described by a single component wavefunction which must decay exponentially inside the potential barrier. Nevertheless, in terms of the density the oscillations average to zero. This phenomenon is known as Zitterbewegung and is associated with relativistic fermions~\cite{Vaishnav2008,Wunderlich2010}.

\begin{figure}[t]
\centering
\subfigure{
\label{fig:ex3-a}
\hspace{4pc} \includegraphics[width=.825\textwidth]{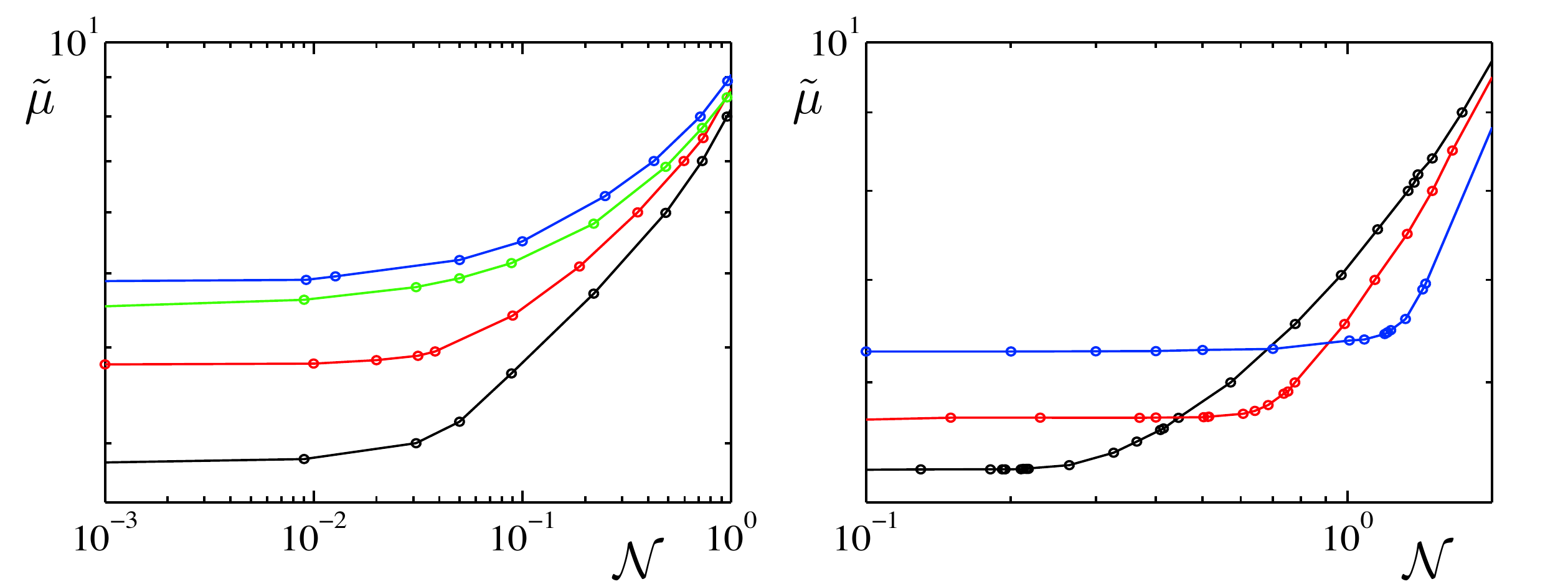}} \\
\caption[]{\emph{Discrete spectra of the dark and bright solitons in a harmonic trap}. (a) Dark soliton spectra. (b) Bright soliton spectra. Single soliton (black), double soliton (red), and triple soliton solution (blue). The vertical axis is labeled by the renormalized chemical potential $\tilde{\mu}$, and the normalization $\mathcal{N}$ is along the horizontal axis. Both quantities are dimensionless. The error bars for each data point are smaller than the point size; curves provide a guide to the eye but do not represent actual data. Note that both axes are in logarithmic scale.         }
\label{DarkBrightSpectra}
\end{figure}

To obtain the functional relation between the chemical potential $\mu$ and the interaction $U$ for a particular excitation inside the harmonic trap, we first derive an expression for the normalization of the wavefunction for the new rescaled NLDE in Eqs.~(\ref{dimensionless3})-(\ref{dimensionless5}), which is found to be
 \begin{eqnarray}
 \int  d\chi  ( |\eta_A(\chi)|^2 + |\eta_B(\chi)|^2) =  \mathcal{N} \, ,  \label{norm3}
\end{eqnarray}
where the right hand side is given by
\begin{eqnarray}
\mathcal{N} = \frac{\sqrt{3}\,  \hbar \omega \, N  \, U}{ 3 \, t_h^2} \, = \,    \frac{\sqrt{3}\,  \omega \, N  \, a^2 \, U}{ 4 \, \hbar \, c_l^2}
\end{eqnarray} 
where $N$ is the number of atoms in the system and we have formulated the expression after the second equality in terms of the lattice constant $a$ and effective speed of light $c_l$. To compute the chemical potential spectra, we fix $\mathcal{Q}$ (which is the same as fixing the relative effects of the lattice geometry and the trap) and vary $\tilde{\mu}$, calculating the norm $\mathcal{N}$ for each value of $\tilde{\mu}$. We thus obtain paired values $(\mathcal{N}, \tilde{\mu})$. These values for the single, double, and triple soliton states are shown in Fig.~\ref{DarkBrightSpectra}.

The plots in Fig.~\ref{DarkBrightSpectra}(a)-(b) show two regimes: weakly nonlinear at small $\mathcal{N}$ versus strongly nonlinear for large $\mathcal{N}$. Note that $\mathcal{N}$ depends on both the total number of atoms and the interaction $U$, as one would expect. For dark solitons in particular shown in Fig.~\ref{DarkBrightSpectra}(a), at small $\mathcal{N}$ ($\sim 10^{-3}$) solutions are weakly nonlinear and correspond closely to the single-particle bound states of a massless Dirac spinor trapped inside a harmonic potential. Here the quantization of spatial modes can be seen by noting that the three lowest multiple dark soliton states intersect the vertical axis at $\tilde{\mu} = 2.83, \, 3.80, \, 4.88$, or in terms of the oscillator frequency $\omega$: $\mu = 2.83 \, \hbar \omega, \,  3.80 \, \hbar \omega , \, 4.88 \, \hbar \omega$. We see that these quantized modes display approximate integer multiples $n$ of the energy $\hbar \omega$: $\mu \approx (2.8 + n) \hbar \omega$. For large $\mathcal{N}$ ($\sim 1$), solutions are strongly nonlinear bound dark solitons with spectra characterized by a power law: $\tilde{\mu} \propto \mathcal{N}^\alpha$. A similar analysis applies to the bright soliton case exhibited in Fig.~\ref{DarkBrightSpectra}(b).

\begin{table}[t]
 \begin{indented}
\item[]\begin{tabular}{@{   \hspace{.5pc}   }llll}                    
 \br 
 \lineup
 Number of   &  Density notch  &  Total        &   Exponent $\alpha$ in   \\   
       dark solitons                    &      depth        &   energy $[\mathrm{nK}]$    &    discrete spectra     \\  
    \mr
  $1$    &           $0.66$          &  $71.95$    &    $0.86$   \\

  2      &         $0.72$              &  $63.57$    &   $0.88$    \\
 
 3  &            $0.77$         &   $53.61$      &   $0.94$   \\ 
     \br 
      Number of   &   Density peak   & Total    &   Exponent $\alpha$ in  \\   
     bright solitons                    &     height           &   energy  $[\mathrm{nK}]$  & discrete spectra     \\  
    \mr
  1        &     $2.86$             &   $100.49$       &   $1.4$      \\

  2         &     $2.78$             &  $112.85$       &   $2.1$     \\

 3         &    $2.63$              &    $131.66$      &    $8.6$       \\
 \br
\end{tabular}   
{\caption{\emph{Visibility of multiple solitons in a harmonic trap.} Density contrasts are measured for the smallest peak in each case. The contrast is computed as the ratios of intensities peak/background, for bright solitons, and notch/background, for dark solitons. Total energies were computed for typical $^{87}$Rb BEC parameters suitable to the NLDE~\cite{haddadcarrsoliton1}. }  \label{table2}}
 \end{indented}
\end{table}

\begin{figure}[t]
\centering
\subfigure{
\label{fig:ex3-a}
\includegraphics[width=.6\textwidth]{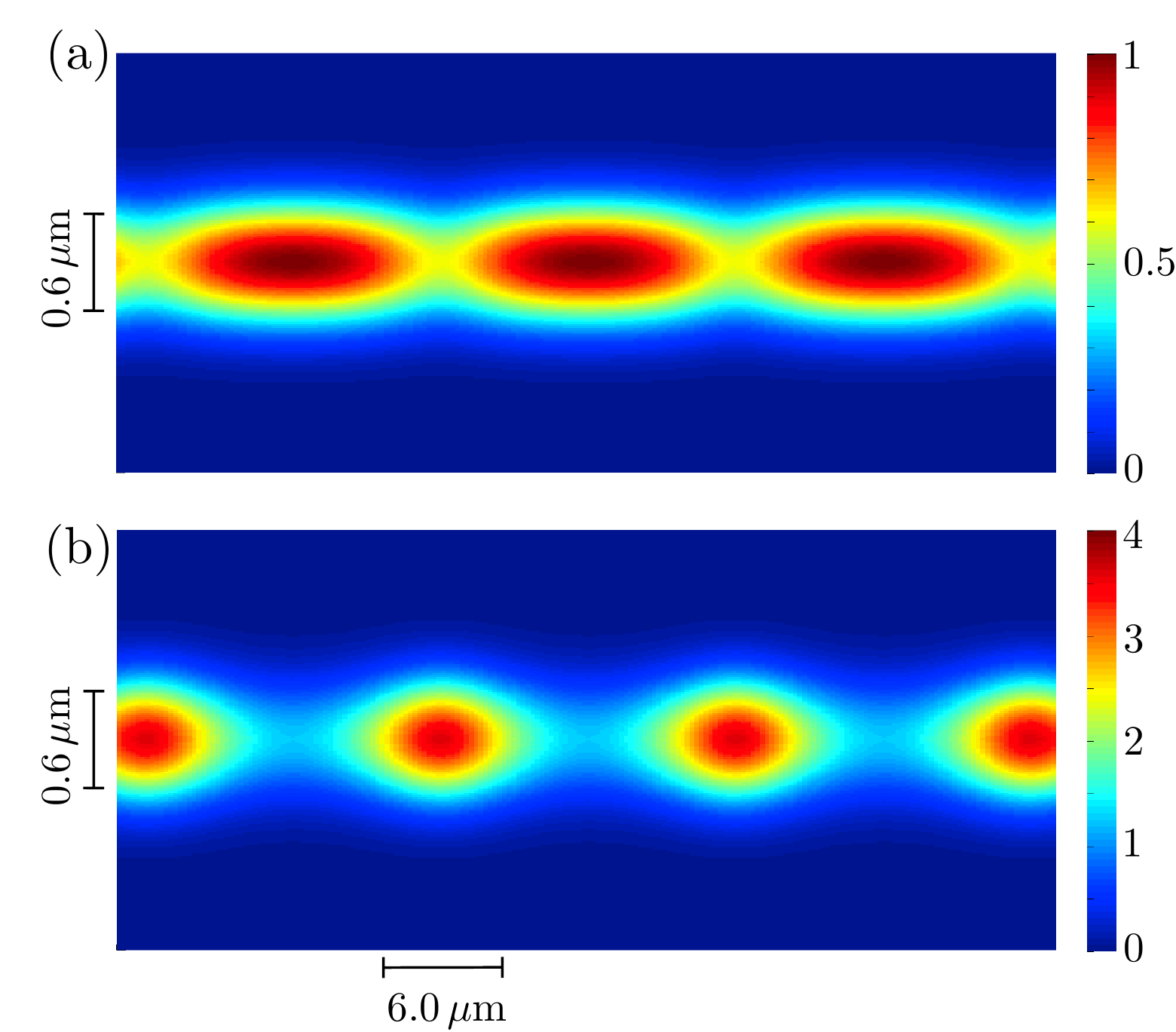}} \\
\caption[]{\emph{Density profiles of multi-soliton states in the absence of the harmonic trap}. (a) Series of four dark solitons. (b) Series of four bright solitons. Note the difference in the horizontal and vertical scales. The length scales correspond to typical parameters for a $^{87}\mathrm{Rb}$ BEC~\cite{haddadcarrsoliton1}.} 
\label{Density2D}
\end{figure}

To obtain values for $\alpha$ in the power law fit $\tilde{\mu} = \beta \, \mathcal{N}^\alpha$, where $\beta$ and $\alpha$ are real constants, we use the Matlab function \emph{polyfit} to obtain a linear fit of the data values $( \mathcal{N}, \, \mathrm{log}_{10} \,   \tilde{\mu}    )$ which returns a two-component vector ${\bf p}$. With ${\bf x}$ the vector of $\mathcal{N}$ values and ${\bf y}$ the vector of $\tilde{\mu}$ values, ${\bf p}  = \mathrm{polyfit}[ {\bf x}, \, \mathrm{log}_{10} ({\bf  y }), \, 1]$ has vector components ${\bf p}(1) = \mathrm{log}_{10}(\alpha)$ and ${\bf p}(2) = \mathrm{log}_{10}(\beta)$, from which we extract the exponent $\alpha = \mathrm{exp}[{\bf p}(1)]$. We find $\alpha =  0.86, \,  0.88, \, 0.94$, for the three lowest multiple dark soliton states, and $\alpha =  1.4, \,   2.1, \,   8.6$, for the lowest bright soliton states. Soliton density profiles and energies are important for visibility in experiments and we list these in Table~\ref{table2}. The density peaks and notch depths for both soliton types were computed for the solitons in Figs.~\ref{DarkSolitonTrap}-\ref{BrightSolitonTrap}. In Fig.~\ref{Density2D} we have plotted the densities for the dark and bright multi-solitons in the zero trap limit. Both solitons extend in a series along the horizontal direction with tight confinement in the vertical direction. Different vertical and horizontal scales are used for ease of viewing. Note also that we use different density color scales in each panel, since dark solitons dip below the asymptotic value of the density, here set to 1 in our units, while bright solitons rise above it.

\section{Conclusion} 
\label{Conclusion}

We have presented soliton stability properties for the quasi-one-dimensional nonlinear Dirac equation and characterized the various excitations in the core and in the bulk. Solitons for both the real and complex Dirac operators are stable with positive or negative real eigenvalues. At finite temperatures, when non-condensate modes are appreciably populated, the negative eigenvalues allow for dissipation into lower energy Bloch states. It is important to note that suppression of these modes at low temperatures is consistent with our interpretation in terms of a metastable background condensate; one sees the same kinds of effects in analogous experiments on non-relativistic dark solitons in BECs described by the nonlinear Schr\"odinger equation~\cite{stellmerS2008}. Finite temperature corrections may be modeled by incorporating a stochastic term or more simply by including a temperature-dependent Bose distribution function to account for finite occupation of higher energy modes. However, we reserve such questions for future investigations.

 We have analyzed the quasi-particle spectrum for modes localized in the soliton core and found an anomalous mode, i.e., a massless Nambu-Goldstone mode associated with phase fluctuations of the core from U(1) symmetry breaking (condensation). Moreover, inside the core we find one zero-energy mode (zero mode) associated with translational symmetry breaking by the soliton. Far from the core the spectrum consists of exotic massive excitations with quartic dispersion in addition to massless Dirac-like excitations. Hence, at low energies and near zero momentum the integrity of the Dirac point is preserved. Moreover, casting our problem in terms of pseudospin degrees of freedom places our results in the context of other domain wall theories. We found that in our case the continuous spectrum far from the core lies in the same universality class as excitations in theories which contain Fermi points such as $^3\mathrm{He}$ and the Standard Model of particle physics~\cite{Volovik2003}.

We have computed the discrete chemical potential spectra for dark and bright solitons bound in a weak harmonic trap. Our results show two clearly distinct asymptotic regions: one for weak nonlinearity where the chemical potential for multiple soliton states differ by a constant multiple of the oscillator energy; and the other limit for strong nonlinearity where the chemical potential obeys a power law. Our numerical solutions confined in the harmonic trap yield ratio values for the notch to bulk contrast in total particle density of $0.66 -  0.77$, for the dark soliton, and $2.63 - 2.86$ for the peak to bulk contrast of the bright soliton. These values were computed for single, double, and triple soliton solutions. In addition, we calculated the range of the total energy for the three lowest multi-soliton states and found these to be $53.61 - 71.95 \, \mathrm{nK}$ and $100.49 - 131.66 \, \mathrm{nK}$ for dark and bright solitons, respectively, for a reasonable experimental parameter set for $^{87}\mathrm{Rb}$ \cite{haddadcarrsoliton1}. Density contrasts and energies offer vital comparative experimental predictions.

\ack{This material is based in part upon work supported by the National Science Foundation under grant number PHY-1067973. L.D.C. thanks the Alexander von Humboldt foundation and the Heidelberg Center for Quantum Dynamics for additional support. We acknowledge useful discussions with Ken O'Hara and Chris Weaver.}

\appendix 

\section{Convergence of numerical solutions of the quasi-1D reduction of the NLDE}

\begin{figure}[phtb]
\centering
 \subfigure{
\label{fig:ex3-a}
\includegraphics[width=\textwidth]{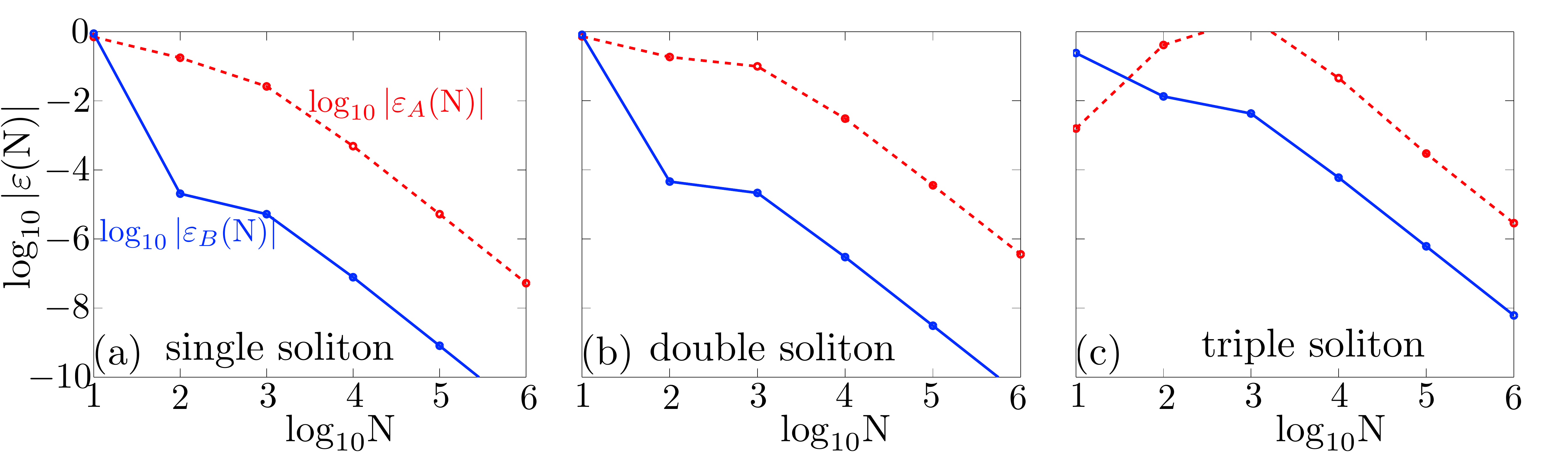}}  \\ 
\caption[]{\emph{Convergence of multiple dark soliton solutions of the NLDE}. Error for the solution depicted in Fig.~\ref{LineSoliton} computed for several values of the grid size for (a) the single soliton, (b) double soliton, and (c) triple soliton. Soliton solutions were obtained by finite differencing using a shooting method to tune the precision of the initial value of $f_A$ near $f_A \approx 1$. Note that the curves are a guide to the eye with data points representing actual data.   } \label{Convergence2}
\end{figure}

To check for convergence of the three lowest soliton solutions depicted in Fig.~\ref{LineSoliton}, the single soliton was obtained by finite differencing using a shooting method to tune the precision of the initial value of $f_A$ near $f_A \approx 1$ with higher precision forcing oscillations to $x$ much greater than the spatial domain of the simulation. The double and triple soliton solutions are then found by tuning the chemical potential $\mu$. For convergence at a single point we compute the solution at $x_i = x/\xi_{\mathrm{Dirac}} = 15$ for several values of the grid size $\mathrm{N} = 10^1, \, 10^2, \, 10^3, \, 10^4, \, 10^5, \, 10^6$ on the same spatial domain size. We use the formula for the error as a function of the number of grid points
\begin{eqnarray}
\varepsilon_{A(B)}(\mathrm{j})  \equiv \left[ \frac{  \psi(x_i)_{A(B)}^{\mathrm{j}+1} - \psi(x_i)_{A(B)}^\mathrm{j})}{ \psi(x_i)_{A(B)}^{\mathrm{j}+1} + \psi(x_i)_{A(B)}^\mathrm{j}) }\right] \, .\label{error1}
\end{eqnarray}
 In Eq.~(\ref{error1}), the subscript $A(B)$ in the symbol $\psi(x_i)_{A(B)}^{\mathrm{j}}$ denotes the sublattice excitation, $x_i$ denotes the $i^{\mathrm{th}}$ element in the discretized spatial coordinate, and the superscript $\mathrm{j}$ denotes the logarithm of the number of grid points used in the calculation, i.e., $\mathrm{j}  =  \mathrm{log}_{10}\mathrm{N}$. In Fig.~\ref{Convergence2} we have plotted $\mathrm{log}_{10}\left| \varepsilon(\mathrm{N}) \right|$ versus $\mathrm{log}_{10}\mathrm{N}$.

\begin{figure}[t!]
\centering
 \subfigure{
\label{fig:ex3-a}
\includegraphics[width=\textwidth]{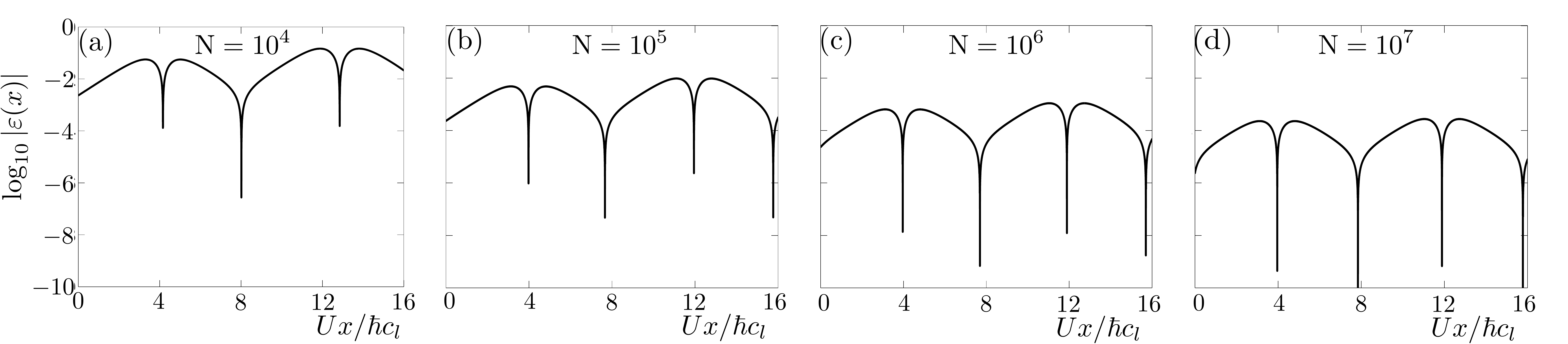}}\\ 
\caption[]{\emph{Convergence of the particle density for the triple dark soliton solution of the NLDE}. Error in the periodicity of the density shown in Fig.~\ref{LineSoliton}(f) for several values of the grid size.} \label{Convergence3}
\end{figure}

Next, we compute the error associated with oscillating solutions by calculating the average difference between values of $\psi_A$ and $\psi_B$ at positions separated by one period. This tells us how the error in the periodicity of our solutions propagates with increasing position. The formula we use for the error $\varepsilon(x)$ is 
\begin{eqnarray}
\hspace{-1pc} \varepsilon_{A(B)}(x)  &\equiv& \frac{ \mathrm{diff}(x) }{ \mathrm{avg}(x) } =  2\left[   \frac{ \psi_{A(B)}( x + L )  - \psi_{A(B)}(x)}{ \psi_{A(B)}( x + L )  + \psi_{A(B)}(x)} \right]  \, , \label{periodicity}
\end{eqnarray}
where $L$ is the periodicity for the particular solution and the error is computed for both two-spinor component functions $\psi_A$ and $\psi_B$. The propagation error in the periodicity of the density shown in Fig.~\ref{LineSoliton}(f) is computed using Eq.~(\ref{periodicity}) with results shown in Fig.~\ref{Convergence3}.

\section*{References}

%\begin{thebibliography}{<1>}

%\bibliographystyle{prsty}

\bibliographystyle{unsrt}

\bibliography{NLDE_Solitons_Refs}

%\end{the bibliography}

\end{document}